\pdfoutput=1
\documentclass[letterpaper,fleqn,usenatbib,useAMS]{mnras}
\usepackage{amssymb,amsmath}
\usepackage{graphicx}
\usepackage{float}
\usepackage[toc,page]{appendix}
\usepackage{color}
\usepackage{multirow}
\usepackage{url}
\usepackage{xcolor}
\usepackage{xspace}
\usepackage{booktabs}

\def\aap{{ A\&A}}

\def\apjl{{ApJL}}

\def\apjs{{ApJS}}

\newcommand{\ie}{\textit{{\it i.e.},~}}
\newcommand{\eg}{\textit{{\it e.g.},~}}
\newcommand{\photoz}{{photo-$z$\xspace\xspace}}
\newcommand{\imshape}{{\textsc{im3shape}}\xspace}
\newcommand{\ngmix}{{\textsc{ngmix}}\xspace}
\newcommand{\bpz}{{\textsc{bpz}}\xspace}
\newcommand{\skynet}{{\textsc{skynet}}\xspace}
\newcommand{\annz}{{\textsc{annz2}}\xspace}
\newcommand{\tpz}{{\textsc{tpz}}\xspace}
\newcommand{\magauto}{{\textsc{mag\_auto}}\xspace}
\newcommand{\msc}{{matched spectroscopic catalogue}}
\newcommand{\wls}{{weak lensing sample}}

\title[{Redshifts of DES Science Verification weak lensing galaxies}]{Redshift distributions of galaxies in the DES Science Verification shear catalogue and implications for weak lensing}

\author[Bonnett, Troxel, Hartley, Amara, Leistedt and the DES Collaboration]{
\parbox{\textwidth}{
\Large
C.~Bonnett$^{1}$, M.~A.~Troxel$^{2}$, W.~Hartley$^{3}$, A.~Amara$^{3}$, B.~Leistedt$^{4}$,
M.~R.~Becker$^{5,6}$, 
G.~M.~Bernstein$^{7}$, 
S.~L.~Bridle$^{2}$, 
C.~Bruderer$^{3}$,
M.~T.~Busha$^{5}$
% ...
\begin{center} (The DES Collaboration) \end{center}}
\\
Affiliations are listed at the end of the paper}

\begin{document}
\date{\today}

\label{firstpage}
\maketitle
\begin{abstract}
We present photometric redshift estimates for galaxies used in the weak lensing analysis of the Dark Energy Survey Science Verification (DES SV) data. Four model- or machine learning-based photometric redshift methods -- \annz, \bpz calibrated against BCC-Ufig simulations, \skynet, and \tpz\ -- are analysed. For training, calibration, and testing of these methods, we construct a catalogue of spectroscopically confirmed galaxies matched against DES SV data. The performance of the methods is evaluated against the matched spectroscopic catalogue, focusing on metrics relevant for weak lensing analyses, with additional validation against COSMOS photo-zs. From the galaxies in the DES SV shear catalogue, which have mean redshift $0.72\pm0.01$ over the range $0.3<z<1.3$, we construct three tomographic bins with means of $z=\{0.45, 0.67, 1.00\}$. These bins each have systematic uncertainties $\delta z \lesssim 0.05$ in the mean of the fiducial \skynet photo-z $n(z)$. We propagate the errors in the redshift distributions through to their impact on cosmological parameters estimated with cosmic shear, and find that they cause shifts in the value of $\sigma_8$ of approx. 3\%. This shift is within the one sigma statistical errors on $\sigma_8$ for the DES SV shear catalog. We further study the potential impact of systematic differences on the critical surface density, $\Sigma_{\mathrm{crit}}$, finding levels of bias safely less than the statistical power of DES SV data. We recommend a final Gaussian prior for the photo-z bias in the mean of $n(z)$ of width $0.05$ for each of the three tomographic bins, and show that this is a sufficient bias model for the corresponding cosmology analysis.
\end{abstract}

\begin{keywords}
cosmology: distance scale -- galaxies: distances and redshifts -- galaxies: statistics -- large scale structure of Universe -- gravitational lensing: weak 
\end{keywords}

\section{Introduction}

One of the key goals of the Dark Energy Survey (DES) is to extract cosmological information from measurements of weak gravitational lensing. Gravitational lensing \cite[for discussion see][and references therein]{1996astro.ph..6001N,2003ARA&A..41..645R,2008PhR...462...67M} involves the deflection of light from distant galaxies by intervening matter along the line-of-sight. Lensing encodes information in the shapes of background objects (\ie galaxies) on both the statistical properties of intervening matter perturbations and cosmological distances to the sources. 
The primary challenge in studying gravitational lensing in the weak regime has been the difficulty in measuring the shapes of galaxies in an unbiased way.
For a detailed discussion of galaxy shape measurements in DES SV, see \cite{jarvisetal2015}. 
However, a weak lensing analysis requires not only the careful measurement of the shapes of galaxies, but also an accurate and unbiased estimate of redshifts to a large ensemble of galaxies. 

Knowing the redshifts of the galaxies in a sample (or equivalently, their distances for a given cosmological model), allows us to differentiate near and distant galaxies and thereby reconstruct the redshift-dependence of the lensing signal.
Hence separating galaxies into redshift bins strongly improves the constraining power of cosmic shear on cosmological model parameters \citep{1999ApJ...522L..21H}. Extensive studies have been reported in the literature that look for optimal configurations of redshift binning and requirements for future ambitious surveys, covering several thousand square degrees, \citep{2007MNRAS.381.1018A,2008MNRAS.386.1219B,2009MNRAS.395.1185C,2009ApJ...699..958S,2010MNRAS.401.1399B,2008MNRAS.387..969A,2012MNRAS.422..553B,2012MNRAS.423..909C,Bordoloi,2012MNRAS.421.1671B,2012ApJS..201...32S,2014MNRAS.444..129C}. 
In addition to gains in statistical precision, separating galaxies into tomographic bins can also mitigate astrophysical systematics. 
For example, moving to a tomographic analysis allows us to better isolate the intrinsic correlations of galaxy shapes in the absence of lensing (see \cite{iareview,2015arXiv150405465K} and references therein), whereas a non-tomographic analysis may otherwise be limited by uncertainties in the impact of this intrinsic galaxy alignment (for more, see \citealt{cosmologypaper}). 

Given the large number of galaxies that make up a lensing sample in a wide field imaging survey, redshifts must be estimated using photometry measured in a series of (typically) broad bands. 
This method of estimating photometric redshifts is known as \photoz \citep[see ][and discussion and references therein]{2010A&A...523A..31H}. Achieving the high level of precision necessary to ensure that the systematic contributions to cosmological parameter uncertainties due to photo-z bias are of the order of the statistical uncertainty is challenging, as is the necessary validation of the derived redshifts 
\citep{2008MNRAS.386..781M,
hildebrandtcfhltlens,
benjamin,
deeplensphotz,
2015MNRAS.446.2523B,
sanchez2014dessva1photoz}. 
Previous weak lensing surveys have tackled this problem in a variety of innovative ways. For example, see \cite{hildebrandtcfhltlens} and \cite{benjamin} for the discussion of this problem in the CFHTLenS survey \citep{erbencfhltlesn, heymanscfhtlens} and \cite{deeplensphotz} in 
the Deep Lens Survey \citep{DLS}.
Substantial and dedicated efforts are required to improve current performance and achieve the target precision in on-going and future surveys. The challenging target set for the full Dark Energy Survey is that the biases in redshift estimates of the means of tomographic bins should be below $\delta z = 0.003$, which is based on the desire to keep redshift systematic errors subdominant to the statistical errors of the lensing surveys \citep{2007MNRAS.381.1018A,2008MNRAS.387..969A}. 

In this work we explore accurate and precisely characterised {\photoz} estimates of $n(z)$, the result of stacking the individual probability distribution functions $p(z)$, with the Science Verification (SV) data of DES. At 139 square degrees, the required precision for DES SV weak lensing analyses are significantly weaker than those for the full DES survey data. As such we target precision at the few percent level for the mean redshifts of a given population of galaxies. This will allow us to have \photoz uncertainties comparable to or lower than the statistical errors on the cosmological parameters we are best able to constrain (\eg $\sigma_8$). We can study the impact of redshift precision directly by propagating the expected \photoz bias to the 
constraints on $\sigma_8$, but also by comparing the differences in final predictions for $\sigma_8$ over the full DES SV shear catalogue from each of four different independent photometric redshift methods.

The paper is organised as follows. In section \ref{sec:data} we introduce the data products that are used in our studies.  In sections \ref{sec:global} and \ref{sec:photoz_glob} we investigate the global properties of the lensing sample including magnitude, colour and redshift distributions. We also discuss the limitations of existing spectroscopic samples. In section \ref{sec:tomo} we extend our analysis to tomographic cases and the impact on cosmological parameters is explored in section \ref{sec:wl}.  Our conclusions are summarised in section \ref{sec:conclusions}.

\section{Data sets}
\label{sec:data}

Prior to the start of the main Dark Energy Survey, the Dark Energy Camera (DECam) \citep{FlaugherDECamStatus2012,DiehlDECam2012,HonscheidControl2012,FlaugherDECamStatus2015}, with a hexagonal footprint of 570 Megapixels, was tested during a preliminary Science Verification (SV) survey from November 2012 to February 2013. These observations produced a useable DES SV galaxy catalogue with which measurement and analysis pipelines have been tested to produce early science results. The DES SV survey mimics full 5-year DES survey parameters over a small patch of the sky, but with significant depth variations due to weather and other challenges during early operations of DECam \citep[see \eg][]{Leistedtetal2015}. The contiguous area used for the DES SV shear catalogue is contained within the South Pole Telescope east (SPT-E) observing region \citep{CarlstromSPT2011}, and covers approximately 139 square degrees in five optical filters, $g$, $r$, $i$, $z$, and $Y$.  We note that the $Y$ band was not used in this work.

In this section we present the DES SV data products relevant for photometric redshift estimation. We also build a catalogue of precise and reliable spectroscopic redshifts by collating a number of proprietary and public spectroscopic datasets that also have DES photometric observations available. This is essential to test the methods for \photoz estimates used in this work. Finally, we describe a set of simulations of the DES SV survey that we use as a secondary method of calibrating and validating the \photoz estimates.

\subsection{DES SV Photometry and Gold Catalogue}

DES data from the SV season were reduced by the SVA1 version of the DES Data Management system \citep{DESDM2012}, using \textsc{SCamp} \citep{BertinSCAMP2006}, \textsc{SWarp} \citep{BertinSWarp2002} and bespoke software packages, as described in \cite{2011arXiv1109.6741S, 2012ApJ...757...83D} and \cite{DESDM2012}. To summarise, the single-epoch images were calibrated, background-subtracted, coadded, and processed in `tiles' ($0.75 \times 0.75$ deg$^2$ squares) defined to cover the entire DES footprint. A catalogue of objects was extracted from the coadded images using \textsc{Source Extractor} \citep[\textsc{SExtractor},][]{BertinSExtractor1996,BertinMorph2011}. In what follows we use $AB$ magnitudes and  \textsc{mag$\_$auto} measurements performed in coadd images, which are reliable for SV galaxies (\eg robust to sharp PSF variations across coadd images) and used in most SV analyses \citep[\eg][]{Crocce2015dessvclustering}.  However, note that shape measurements are performed in single-epoch images with a dedicated pipeline using multi-epoch fitting techniques, as described in \cite{jarvisetal2015}. The analysis presented in this work will be concerned with the objects that meet the quality cuts of that pipeline.

The main catalogue of reliable objects in DES SV is the Gold catalogue described in \cite{Rykoffetal2015}. It starts with all objects detected in SV images and successively applies quality cuts to reject objects and regions that are deemed problematic (\eg regions with poor observations or photometry). To be included in the Gold catalogue, a galaxy must: 

\begin{description}
  \item [$\cdot$]be observed at least once in all four $griz$ bands,
  \item [$\cdot$]be at a declination above $-61^{o}$ to avoid regions of bad photometric calibration (\eg Large Magellanic Cloud)
  \item  [$\cdot$]not be in regions with galaxy surface density $>3\sigma$ below the mean.
  \item [$\cdot$]not be in regions surrounding bright stars.
  \item [$\cdot$]not be in regions with a concentration of large centroid shifts or dropouts between bandpasses.
\end{description}
Further information on star-galaxy separation and quality cuts at the shape measurement level are described in detail in \cite{jarvisetal2015}.

\subsection{DES SV Shear Catalogue}

\begin{figure}
\begin{center}
\includegraphics[width=\columnwidth]{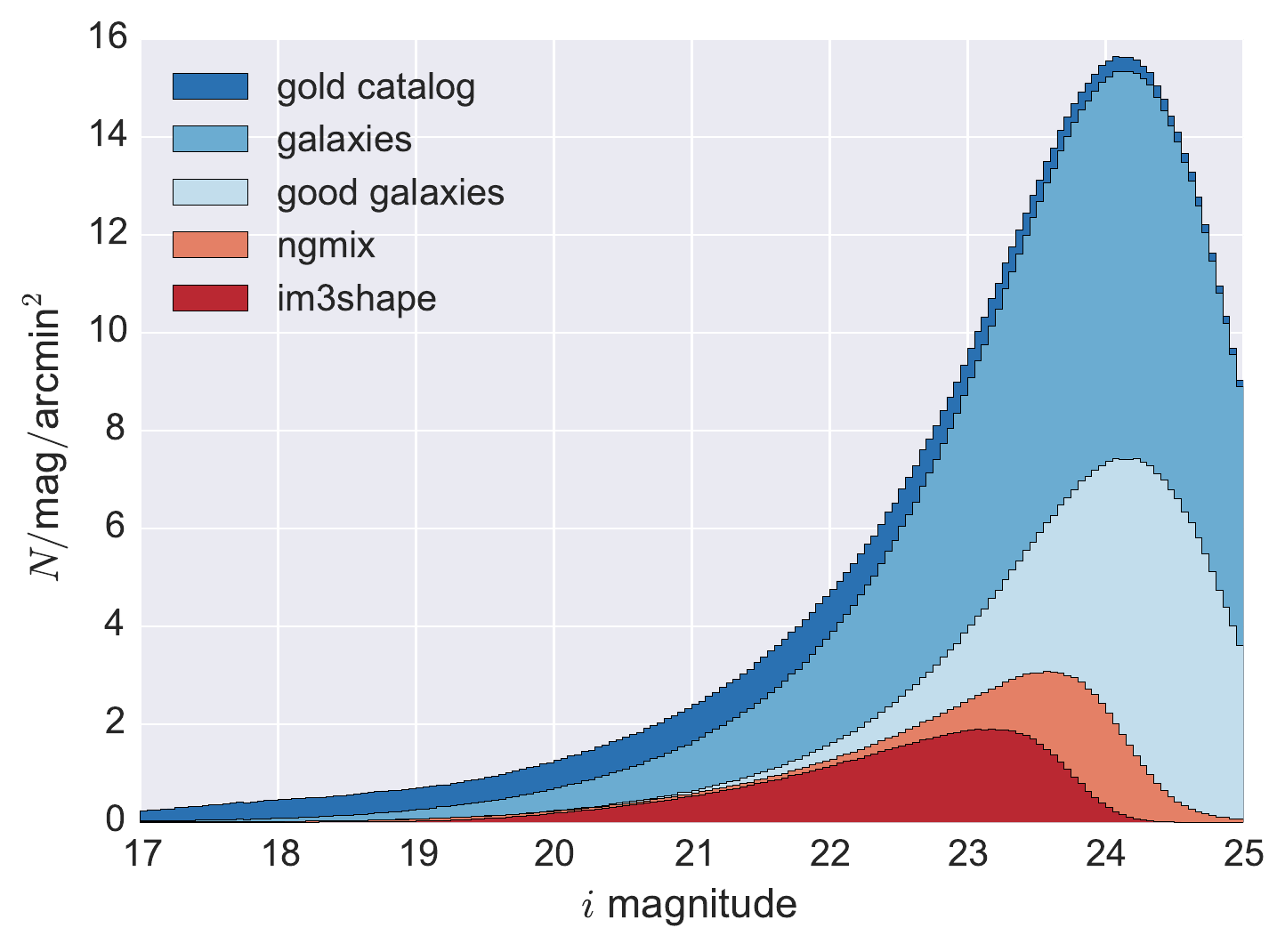}
\end{center}
\caption[]{$i$-band magnitude histograms for various levels of cuts from the full Gold catalogue down to the final shear catalogue. \label{fig:maghist}
}
\end{figure}

Two semi-independent shear pipelines -- \imshape\ and \ngmix\ -- have been produced for a subset of objects in the DES SV Gold Catalogue in the SPT-E region of the sky. These are described further in \cite{jarvisetal2015}, but relevant details are summarised below. The two shear pipelines produce separate shear measurements for each galaxy, and thus select a different subset of the galaxies in the Gold Catalog as having well-measured shears. This leads to a different population of galaxies used by either pipeline in constructing the $n(z)$ for each tomographic bin in a weak lensing analysis, though the \imshape selection is nearly a subset of the \ngmix selection. The final shear catalogue is the intersection of the gold galaxy selection, these shear-related cuts, and a final 'good' galaxy selection for lensing that removes objects with \textsc{SExtractor} flags $=1,2$, very lower surface brightness objects, very small objects, or those with colours outside reasonable bounds ($-1<g-r<4$ and $-1<i-z<4$). These selection effects also produce slightly different photometric properties in the galaxy sample used. This is demonstrated in Fig. \ref{fig:maghist}, where the $i$-mag histogram is compared for all `Gold' objects, all galaxies, `good' galaxies, as defined above, and finally the two shear selections.

\begin{itemize}

  \item {\bf \imshape} The \imshape\ shear measurement pipeline is built on the \imshape\ code discussed in \cite{zuntzetal2013} and modified as described in \cite{jarvisetal2015}. The \imshape\ code is a forward-modelling maximum-likelihood method that fits two galaxy models to an image in the $r$ band: an exponential disc and a de Vaucouleurs bulge. The best-fitting model is then used to estimate the ellipticity. Inverse variance weights are calculated for each galaxy empirically in bins of size and signal-to-noise. The final \imshape\ shear catalogue has a number density of $\simeq$4.2 galaxies per square arcminute.
  
\item {\bf \ngmix} The \ngmix\ shear measurement pipeline represents simple galaxy models as the sum of Gaussians \citep{hogglang2013}. The same model shape is fit simultaneously across the $riz$ bands, with parameters sampled via Markov Chain Monte Carlo (MCMC) techniques. Ellipticities are then estimated using the \textsc{lensfit} algorithm \citep{miller2007} with priors on the intrinsic ellipticity distribution from \textsc{Great3} \citep{great3}. Inverse variance weights are calculated for each galaxy from the covariance of the shape estimate and an intrinsic shape noise estimate. The final \ngmix\ shear catalogue has a number density of $\simeq$6.9 galaxies per square arcminute.

Throughout this work we use the \ngmix\ catalogue as the default \wls\ unless explicitly stated otherwise. 

\end{itemize}

\subsection{Spectroscopic Catalogues}
\begin{figure}
\begin{center}
\includegraphics[width=\columnwidth]{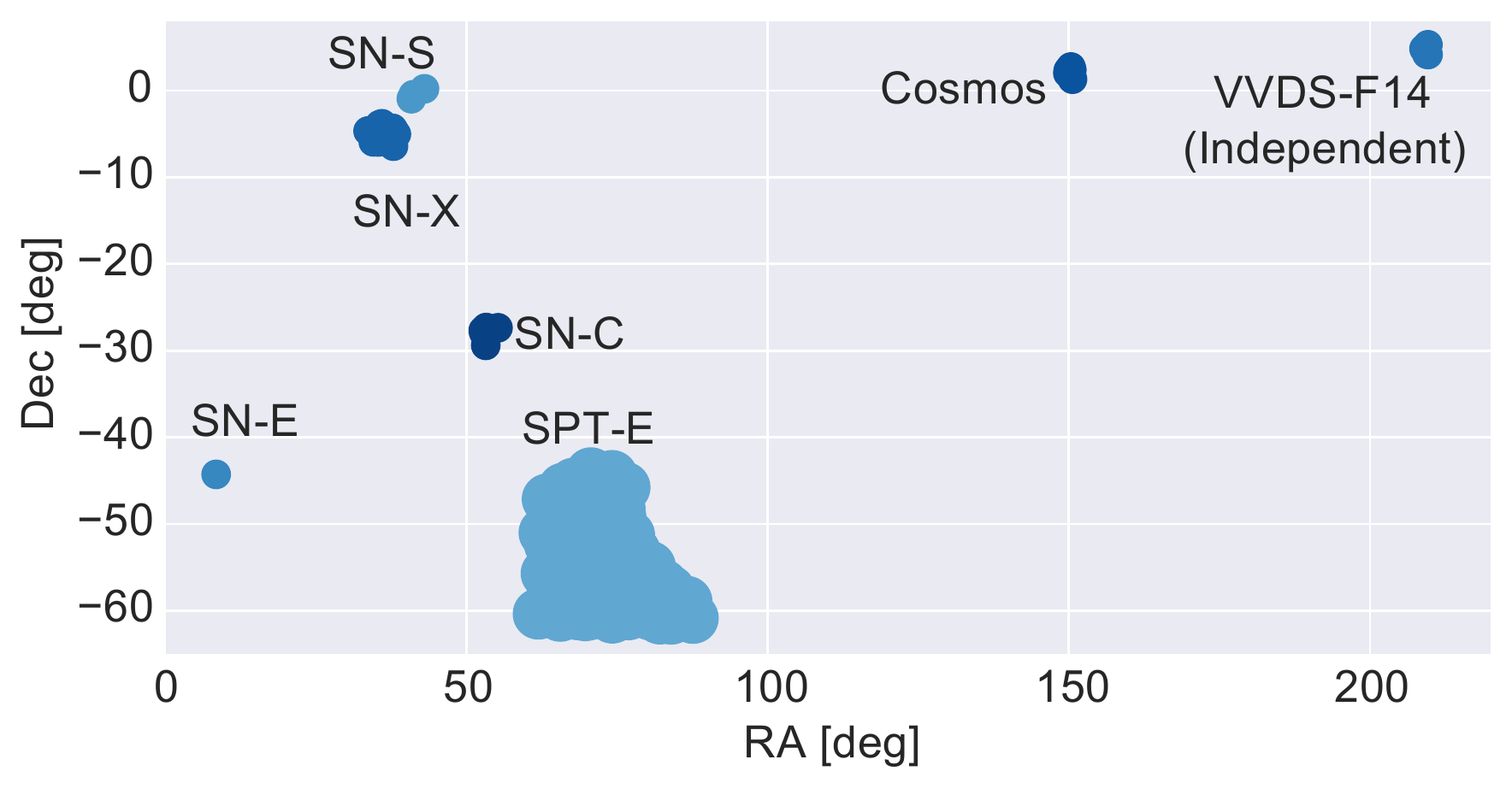}
\caption{Location of the six spectral fields and the main DES SV (SPT-East) field on the sky. The SN fields are the DES supernova fields while the other two have been observed with DECam outside of the DES survey.}
\label{fig:sky}
\end{center}
\end{figure}

To train and assess the performance of the photometric redshifts we assemble a matched catalogue of 
galaxies that are observed with both DECam and a spectrograph.
In this section we describe the photometric and spectroscopic properties of this matched catalogue. Objects are matched on the sky within a matching radius of 1.5 arcseconds. The spectra used come from 6 distinct areas on the sky and contain a total of 46139  galaxies. The distributions of these fields on the sky relative to the main DES SV SPT-E field are shown in Fig. \ref{fig:sky}. 
In Table \ref{table:spec} the general properties of the spectroscopic surveys used in this matched catalogue are listed, but for a more detailed description
of the properties (\eg the quality flags used), we refer the reader to Appendix \ref{section:app_a}.

\begin{table}
\begin{tabular}{lrrr}
\toprule
             Spectroscopic survey &  Count &  Mean $i$ &  Mean z \\
\midrule         
                    VIPERS &   7286 &          21.52 &    0.69 \\
                      GAMA &   7276 &          18.61 &    0.22 \\
                   Zcosmos &   5442 &          20.93 &    0.51 \\
             VVDS F02 Deep &   4381 &          22.40 &    0.68 \\
                      SDSS &   4140 &          18.82 &    0.39 \\
                      ACES &   3677 &          21.73 &    0.58 \\
                  VVDS F14 &   3603 &          20.61 &    0.49 \\
               OzDES &   3573 &          19.85 &    0.47 \\
                ELG cosmos &   1278 &          22.22 &    1.08 \\
                      SNLS &    857 &          21.09 &    0.55 \\
                 UDS VIMOS &    774 &          22.54 &    0.85 \\
                    2dFGRS &    725 &          17.52 &    0.13 \\
                     ATLAS &    722 &          18.96 &    0.35 \\
           VVDS spF10 WIDE &    661 &          21.16 &    0.53 \\
           VVDS CDFS DEEP  &    544 &          22.05 &    0.62 \\
                 UDS FORS2 &    311 &          23.80 &    1.25 \\
             PanSTARRS MMT &    297 &          19.94 &    0.35 \\
           VVDS Ultra DEEP &    264 &          23.71 &    0.88 \\
         PanSTARRS AAOmega &    239 &          19.69 &    0.32 \\
              SNLS AAOmega &     81 &          21.16 &    0.56 \\
\bottomrule
\end{tabular}
\caption{The number of galaxies that are included in the matched spectroscopic catalogue are listed for each spectroscopic survey with the corresponding mean redshift and mean $i$ band magnitude.
Further details can be found in  appendix \ref{section:app_a}.
\label{table:spec}}
\end{table}

\begin{figure}
\begin{center}
\includegraphics[width=\columnwidth]{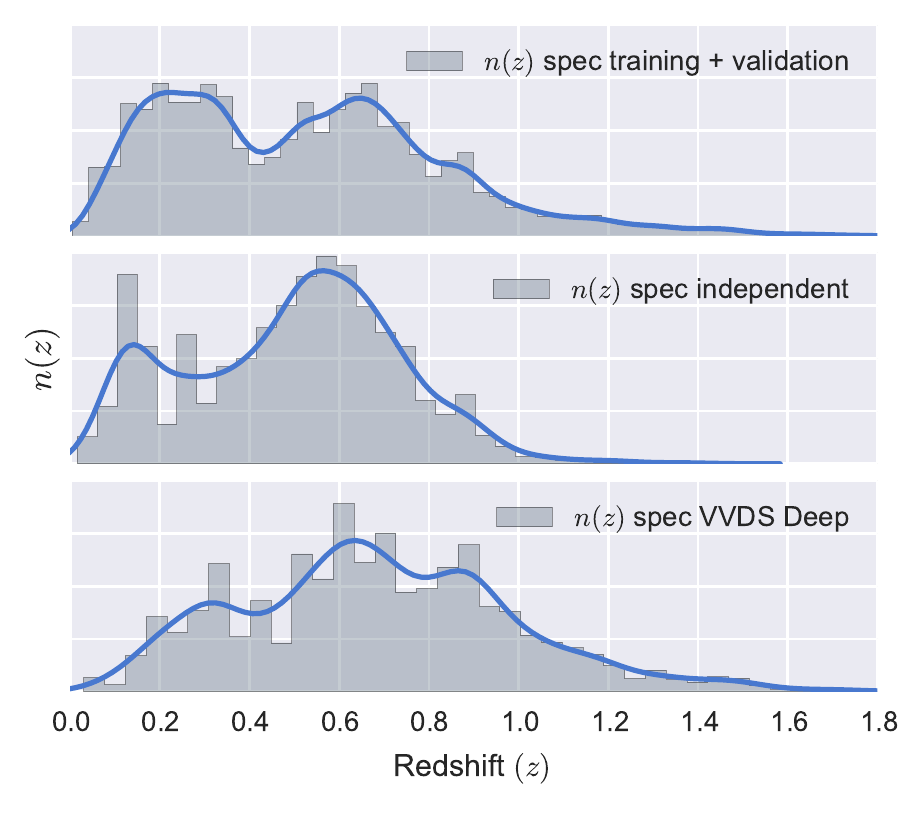}
\end{center}
\caption[]{The normalised redshift distributions of the spectroscopic samples used in producing and testing the photometric redshift estimates. The solid line is the Kernel Density Estimate (KDE) \citep{ML} estimate of the underlying density. Top panel: The combined training and validation samples. Middle panel: The independent sample (VVDS-F14). Bottom panel: The VVDS-Deep sample.
\label{fig:spec-z-dist}}
\end{figure}

The final matched spectroscopic catalogue has been cleaned of objects that we do not expect to be present in the shear catalogue. This includes removing all stars, strong lenses, and AGN. The matching is limited to the $(0 < z < 1.8)$ redshift range.
This means that for all the machine learning (ML) methods used in this work the density of $n(z)$ above $z=1.8$ will be zero, though model fitting codes do not have this drawback.
We test that artificially cutting the $n(z)$ at 1.8 for a model fitting code biases
the constraints on $\sigma_8 $ at the 1\% level, which is sufficiently small relative to the statistical error (see Sec. \ref{sec:wl} for more details).

We divide the resulting matched spectroscopic catalogue into three samples: a training, a validation, and an independent sample, which are compared in Fig. \ref{fig:spec-z-dist}.
The independent sample contains all the matched galaxies from VVDS-F14 field; a total of 3,603 galaxies. 
This field is spatially removed from the other spectroscopic fields, as shown in Fig. \ref{fig:sky}, and therefore the line of sight structure within this field is uncorrelated with that of training and validation sets. 
The use of this field will allow us to assess issues pertaining to sample variance and radial learning in the machine learning methods (\eg App. \ref{app:primus}).
If the redshift solution is overtrained or subject to systematic incompleteness, any performance metrics on a validation set with a near identical redshift distribution to the training sample would be too optimistic.
In App. \ref{app:primus}, we demonstrate an example of extreme selection effects in a training set based on the PRIMUS survey, while in Sec. \ref{sec:spec_compl} we study the completeness of the training set used in this work. 
The remaining 42,536 galaxies in the matched spectroscopic catalogue are split into the training and validation samples containing, respectively, 70\% and 30\% of the galaxies. This retains a total of 28,219 galaxies in the training sample and 14,317 galaxies in the validation sample.

\subsection{COSMOS Data}
\label{sec:cosmosdata}

In addition to spectroscopic data from the literature, we also make use of the point-estimated photometric redshifts from \cite{ilbert2009} in the COSMOS field. These \photoz estimates were computed from 30-band photometry 
with the Le Phare template fitting photometric redshift code \citep{lephare}. 
The COSMOS field was observed with DECam during the SV observing period and coadd images with a similar total exposure time as the SV survey have been produced.
We match the catalogue extracted from these images to the COSMOS \photoz sample, and trim to a subsample representative of the shear catalogue. This trimming was performed by applying cuts in the $i$-band FWHM - magnitude plane as follows:

\begin{align*}
{\rm FWHM~(arcsec)} &> 0.105 \times i~({\rm mag}) - 1\\
{\rm FWHM~(arcsec)} &> 0.751 \times i~({\rm mag}) - 15.63\\
i &> 18~({\rm mag})
\end{align*}

\noindent together with a surface brightness cut at $\mu_{\rm eff}~<~28~{\rm mag~arcsec}^{-2}$. These cuts approximate the final shape catalogue selection function and allow us a further independent estimate of the redshift distribution of the weak lensing sample.

\subsection{Simulated SV data: the BCC-UFig}
\label{sec:bccufig}

In the following sections we will calibrate a model based photo-z method using a set of galaxy catalogues extracted from simulated SV data: the BCC-UFig \citep{Chang2014bccufig}. The latter is based on simulated DES coadd images created using the Ultra-Fast Image Simulator \citep[UFig, ][]{Berge2013ufig}. The input galaxy catalogues for these images were taken from the Blind Cosmology Challenge \citep[BCC, ][]{busha2013bcc}. The galaxy catalogues were then obtained by running source extraction and processing codes to mimic the pipeline run on the real DES SV data, as described in \cite{Chang2014bccufig} and \cite{Leistedtetal2015}. The BCC-UFig was shown to reliably mimic the SV data in terms of colour, redshift, and spatial distributions of the objects, and also reproduce systematics observed in the reduced galaxy catalogues such as spatially varying depth and correlations with observing conditions \citep{Chang2014bccufig, Leistedtetal2015}. In this paper we push the comparison further and consider catalogues similar to the weak lensing catalogue described above by making the same catalogue-level cuts as are used for the COSMOS data.

\section{Properties of matched spectroscopic catalogue and templates}
\label{sec:global}

Ideally, we would be able to compile a sample of spectroscopically identified objects that are fully representative of our target weak lensing galaxy population. If these spectroscopic objects were sufficiently numerous and well-sampled over the sky, then the redshift distribution of these objects could be used in conjunction with weak lensing measurements to infer constraints on cosmological parameters. However, even in large samples such as the one compiled for this work, biases remain due to spectroscopic incompleteness and difficulties in representing all galaxies in the face of spatially varying data quality. 

In this section we investigate to what extent our existing spectroscopic sample should reflect the underlying redshift distribution of our photometric sample and assess the effectiveness of weighting spectroscopic objects in correcting for differences between the photometric and spectroscopic galaxy populations.
We pay special attention to possible biases in the inferred probability distribution of the weak lensing sources due to these limitations."
Note that while modelling methods do not require representative training samples, biases may still arise if the model templates are not a sufficiently accurate description of the data.
This is analogue to model bias in cosmic shear measurements \citep{2010MNRAS.404..458V,2014MNRAS.441.2528K}. 
As in cosmic shear, we can aim to tackle these issues through simulations of the data. Thus Secs. \ref{section:noise}-\ref{sec:spec_compl} address challenges related to machine learning methods, while Sec. \ref{section:templates} discusses challenges to using template fitting methods.

\subsection{Noise properties of the matched catalogue}\label{section:noise}

\begin{figure*}
\begin{center}
\includegraphics[scale=0.80]{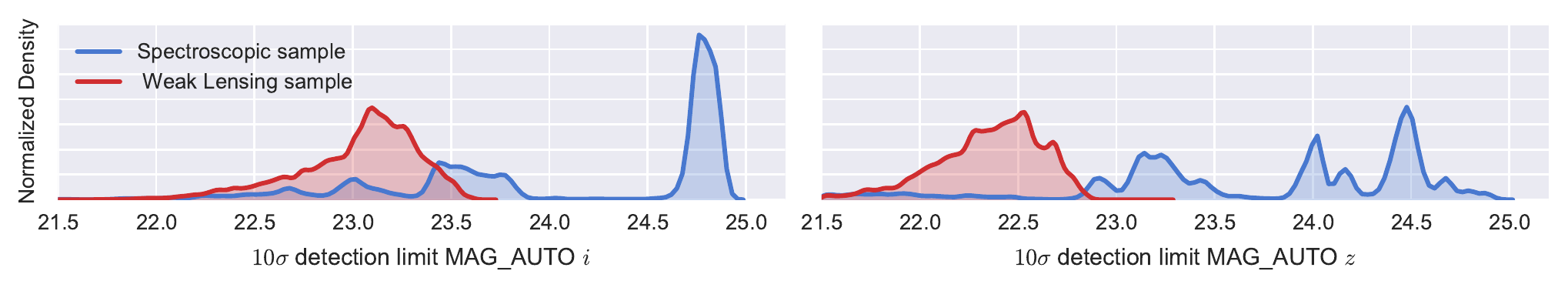}
\end{center}
\caption{The $10 \sigma$ \magauto\ detection limits of the matched spectroscopic sample (blue) compared to that of the weak lensing sample (red). 
The \msc\ has a significantly larger detection limit due to the fact that many DES galaxies with spectra lie in the frequently observed DES supernova fields.
} 
\label{fig:depth_plot}
\end{figure*}

A large fraction of the DES-SV galaxies that have spectra lie in the DES supernovae fields or other fields with a significantly longer cumulative exposure time than the SPT-E field, which contains the galaxies used for the weak lensing science. 
We show in Fig. \ref{fig:depth_plot} the estimated $10 \sigma $ \magauto\ detection limits of the \msc\ compared to that of the \wls.
The $10 \sigma$ detection limits differ significantly between the samples, with the galaxies in the matched spectroscopic catalogue having significantly deeper detection limits on average.
This poses a problem for ML methods as they do not explicitly take the noise measurement into account.
The ML methods in this work implicitly assume the noise properties from the \msc\ to be identical to those of the \wls.

One way to obtain a similar depth distribution in the spectroscopic set is to create co-added images of the deeper fields using a subset of exposures with numbers similar to those typical in the SPT-E field, as was used in \cite{sanchez2014dessva1photoz}.
A second option is to algorithmically degrade the photometry of the \msc\ for the bands of the galaxies with higher $S/N$. 
This is done in the following manner: 
\begin{enumerate}
\item For every galaxy in the \msc, we find its nearest neighbour in four-dimensional colour-magnitude space from the \wls. 
\item If one or more bands of the matched galaxy have a fainter $10 \sigma $ detection limit than the \wls\ detection limit in those bands, then a new magnitude is drawn. 
\item This new magnitude is determined according to a normal distribution using the measured magnitude of the spectroscopic galaxy as the mean and the error on the magnitude of the selected neighbour in the \wls\ for the variance.
\end{enumerate}

The limits in image depth ($10 \sigma$ detection) for which we decide to re-draw a new magnitude value are: \magauto\ $g = 24.5$, \magauto\ $r = 24.3$, \magauto\ $i = 23.5$, and  \magauto\ $z = 22.8$.
So, for a galaxy in the \msc\ that has a $10 \sigma$ detection of 24.7 in $i$ band and a  $10 \sigma$ detection of 22.5 in $z$ band we draw a new $i$ band magnitude and keep the original $z$ band magnitude.

This leads to a matched spectroscopic catalogue that has approximately the same noise properties as the weak lensing sample.
The method has some advantages over re-stacking, one of which is that we can degrade to any other noise level as long as the original exposures are of sufficient depth. 
This is not necessarily possible with re-stacking due to the fact that observing conditions sampled during pointings in SPT-E cannot be recreated with those observed in the deeper fields.   
To protect against potential biases introduced by this procedure, the training and validation in this work have been algorithmically degraded while the independent field containing all the VVDS-F14 galaxies is created by re-stacking and is identical to the reduction of the field used in \cite{sanchez2014dessva1photoz}.
We validated that using restacked cooads instead of resampling the magnitudes has no significat effect on our results.

\subsection{Weighting of the spectroscopic set}\label{sec:weights}

\begin{figure}
\begin{center}
\includegraphics[width=\columnwidth]{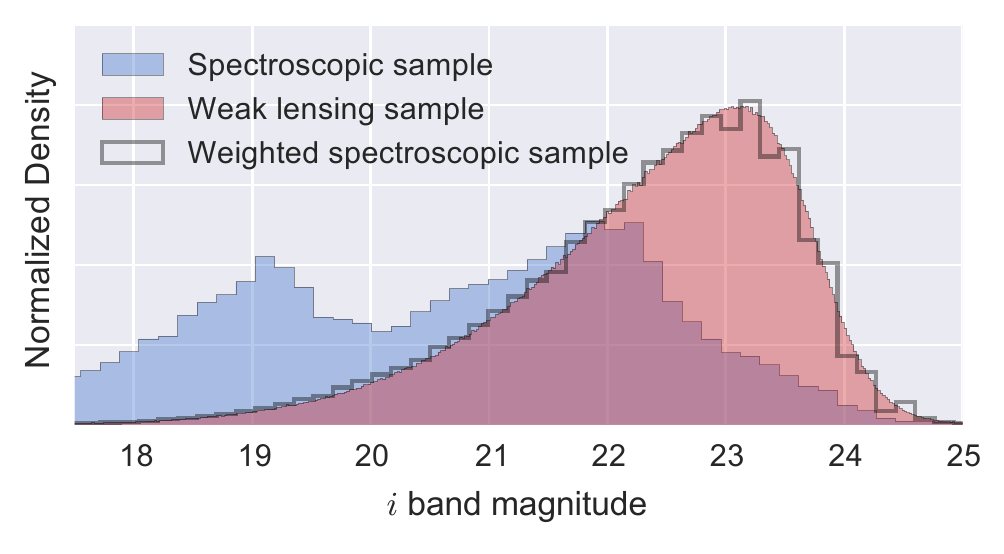}
\end{center}
\caption[]{The $i$-band magnitude distribution of the \msc\ in shown in blue and the \wls\ is shown in red. The \msc\ after weighting is shown as the grey histogram outline overlaying the \wls. \label{fig:specphotsamples}}
\end{figure}

\begin{figure*}
\begin{center}
\includegraphics[scale=0.80]{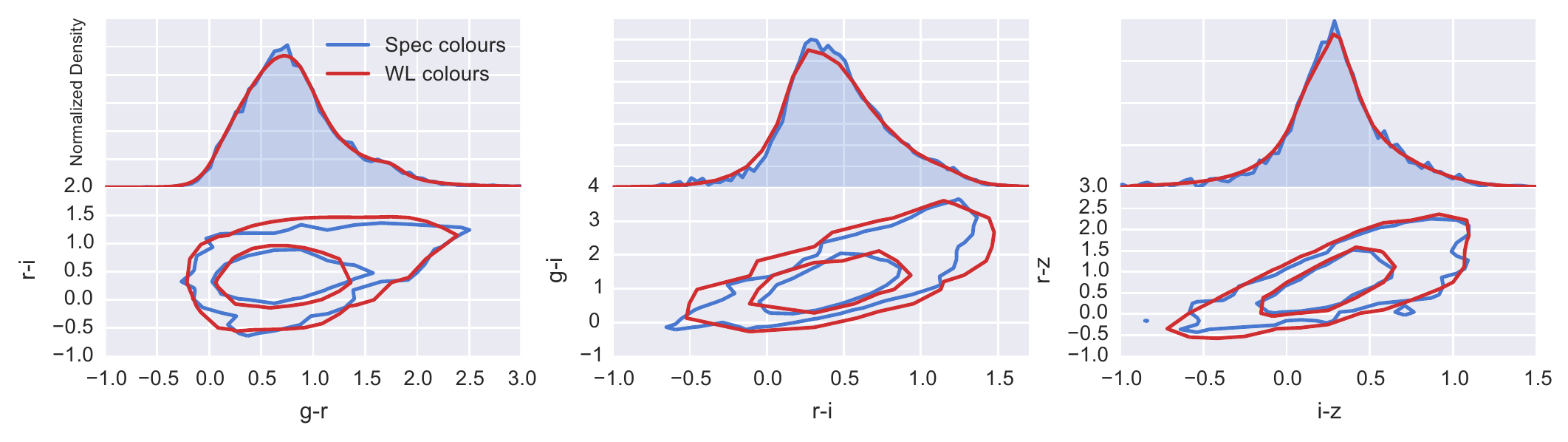}
\end{center}
\caption{The colour distribution of the weighted \msc\ is shown in blue relative to the \wls\ in red. Top row: 1D histograms of the three colours: $g-r$, $r-i$, and $i-z$. 
Bottom row: Related 2D comparisons of the colour distributions. In general the weighted \msc\ colour distribution matches the \wls\ colour space well. 
although in the bottom row we can see that weighted \msc\ is unable to match the tails of the \wls.
\label{fig:colour2}}
\end{figure*}

In the work presented here we characterise the impact of errors in redshift estimation on weak lensing studies. Our focus is thus on the galaxy samples selected based on our ability to measure accurately their shapes in DES SV.  Figure \ref{fig:specphotsamples} shows the $i$-band magnitude distribution of the \msc\  in blue and the distribution of the weak lensing sample from DES SV in red. The difference in magnitude of the samples is very clear, with the matched spectroscopic sample biased to brighter magnitudes.
We account for differences in magnitude and colour by weighting galaxies in the spectroscopic sample in such a way that the weighted distribution of training galaxies matches the weak lensing source distribution.
This can then be used in performance metrics to give a better indication of the likely errors coming from averaging over the weak lensing population. The weights we use are calculated as in \cite{sanchez2014dessva1photoz} by estimating the density of objects in the matched spectroscopic sample in colour-magnitude space noted below, with all objects detected in all bands, and:
\begin{align*}
-1 < g&-r < 4\\
-1 < r&-i < 4\\
-1 < i&-z < 4\\
16 &< i\\
16 &< r.\\
\end{align*}
We then compare this density with the density of the \wls\ at the same location in colour-magnitude space, using the \ngmix\ catalogue.
The  ratio of the densities of the \wls\ to the \msc\ at the location of a spectroscopic galaxy in colour-magnitude space is calculated by counting the number of galaxies in the \wls\ in a hypersphere with radius to the 5$^{th}$ nearest neighbour in Euclidian space in the \msc. 
The normalised ratio of these densities are then used as weights for the spectroscopic galaxies (see \citealt{lima} for more details on the implementation). 
 
Fig. \ref{fig:specphotsamples} shows the weighted $i$-band distribution for the spectroscopic sample, which better matches the \ngmix catalogue.
In Fig. \ref{fig:colour2}, we show $g-r$, $r-i$, and $i-z$ for the \msc\ and \wls\ on the top row while we show $g-r$ vs $r-i$, $r-i$ vs $g-i$ and $i-z$ vs $r-z$ in the bottom row. The weighted colours of the \msc\ are a good match to those of the \wls, although we can see in middle panel of the bottom row that the tails of the colour distributions of the \wls\ are  not as well approximated.
This is due to the fact the  \msc\ only has $\thicksim $40,000 galaxies while the \wls\ has more than 3,000,000, hence the tails of the distributions of the \wls\ are poorly sampled by the limited amount of objects in the \msc.

We find that 1.6\% of the \wls\ fall outside the range of colours sampled by our spectroscopic catalogues.
It is relatively straightforward to remove these regions, but the results in this work are robust to the inclusion or exclusion of these 1.6\% galaxies.


\subsection{Assessing the weighted spectroscopic sample}\label{sec:spec_compl}

The weighting procedure assumes that small regions of colour-magnitude space (pixels) populated by galaxies in the \wls\ are fairly sampled in the \msc.
If this is the case, then weighted estimates of performance metrics will be equivalent to those obtained from a complete spectroscopic sample (\ie one without biases due to a selection function or incompleteness).
However, it is possible that some galaxies live in colour-magnitude regions where incompleteness could lead to missing populations from the spectroscopic sample. 
The redshifts of the spectroscopic sample in these regions could then be biased relative to the full sample of DES galaxies that lie in the same regions of colour space.

\begin{figure}
\begin{center}
\includegraphics[width=\columnwidth]{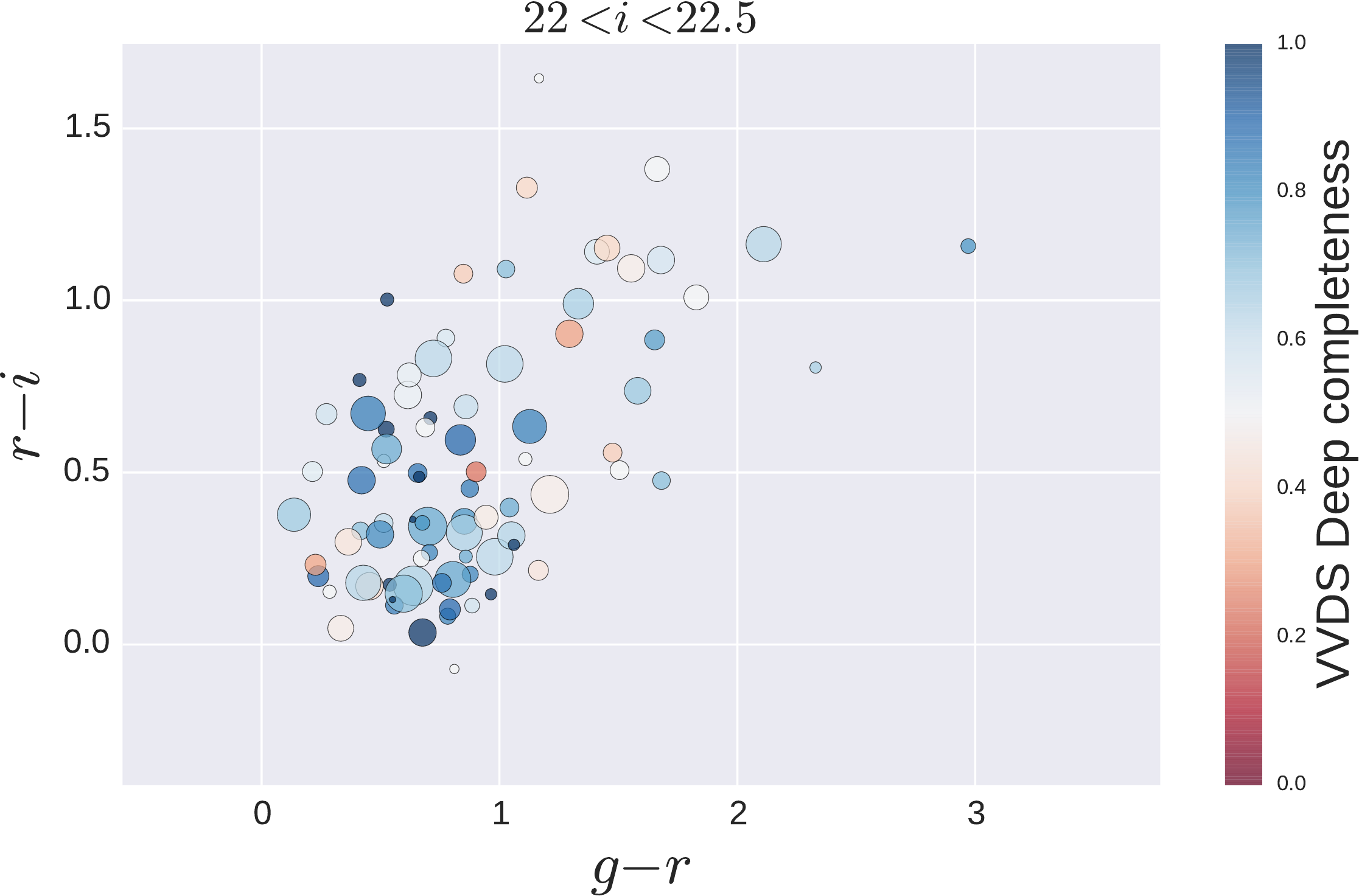}
\includegraphics[width=\columnwidth]{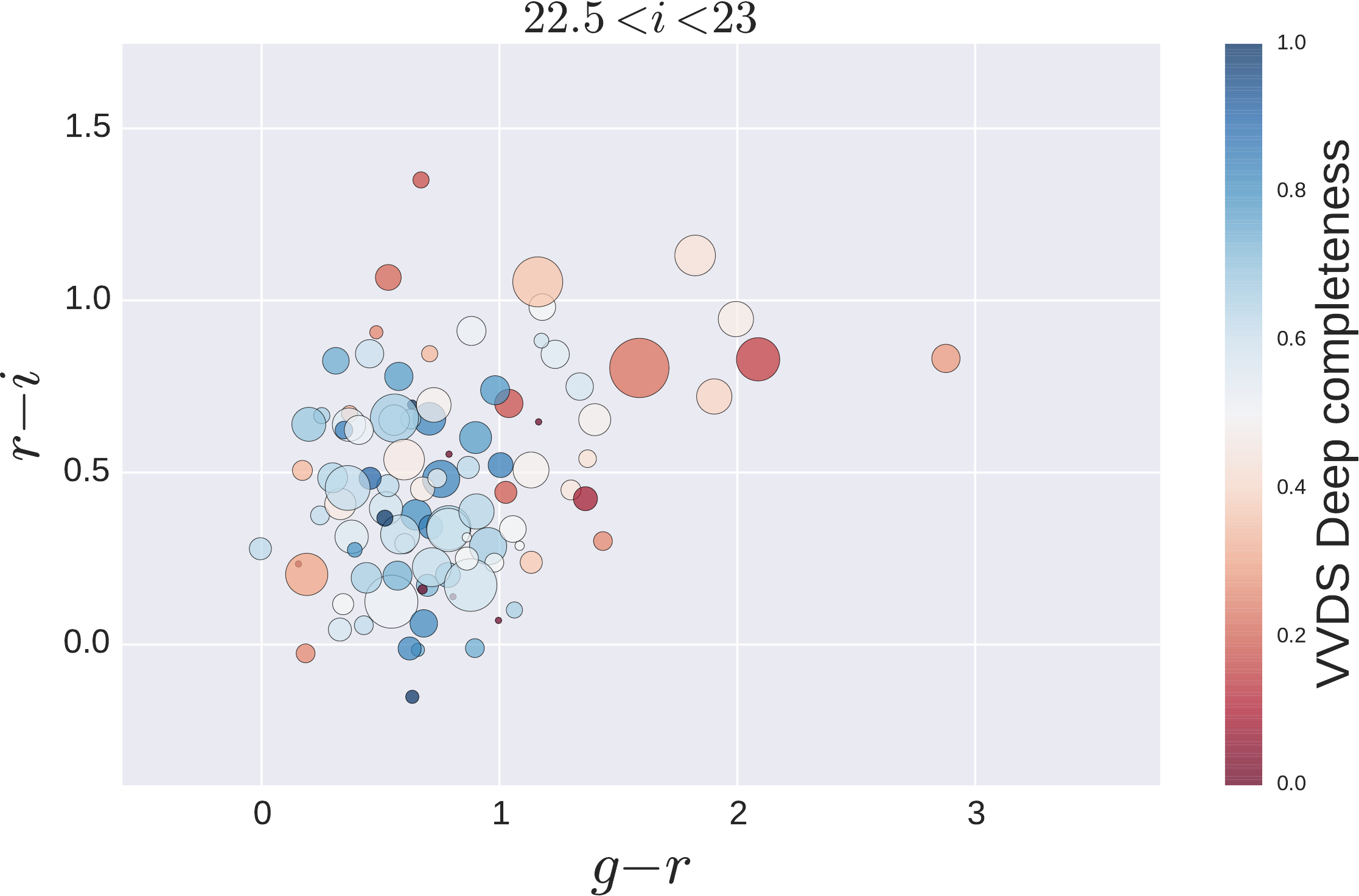}
\includegraphics[width=\columnwidth]{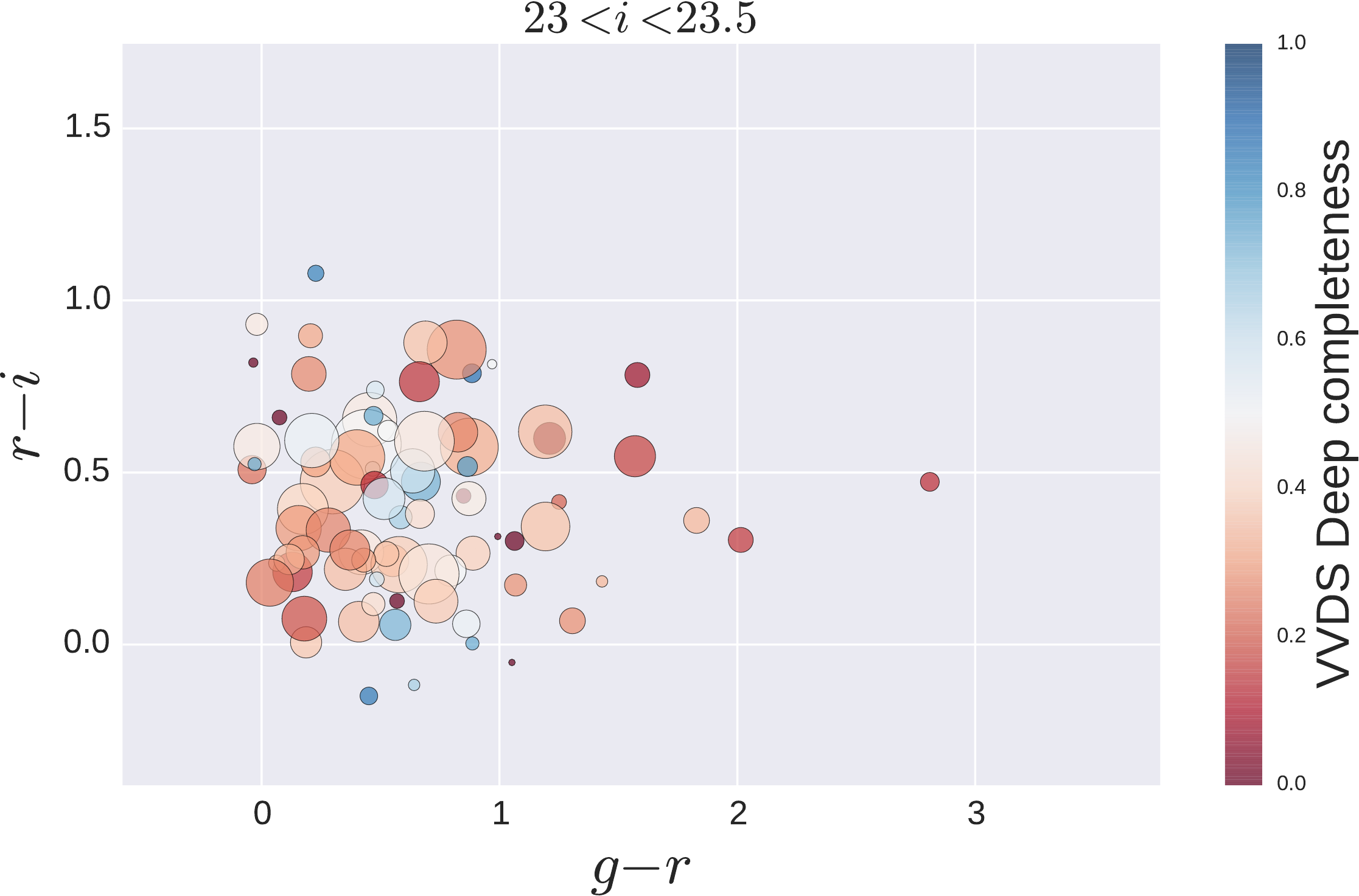}
\end{center}
\caption{Spectroscopic completeness of the VVDS Deep sample in $g-r$ vs $r-i$ colour space. Each point represents the centre of a 4-D colour-magnitude k-means cell containing a similar number of galaxies from the DES SV \ngmix\ catalogue.
The size of the point represents the number of targeted objects, while the colour indicates the fraction that returned a reliable redshift. The three magnitude ranges (as labelled) cover the $i$-band magnitude range that contains the majority of galaxies in the weak lensing sample -- see Fig. \ref{fig:maghist} for the distribution in the catalogues. 
\label{spec_incompl}}
\end{figure}

The only sizeable sample that we have access to with target selection of comparable depth to DES is the VVDS Deep survey. This sub-survey within VVDS targeted galaxies purely on $i$-band magnitude at $i<24$.  
In order to understand how the incompleteness within this survey corresponds to the colour and brightness of the galaxy distribution we break the four-dimensional colour-magnitude volume of the weak lensing sample ($g-r$, $r-i$, $i-z$ and $i$-band magnitude) into cells based on a k-means clustering algorithm \citep{ML}. Each cell represents approximately $0.2\%$ of the sample. To each of these k-means cells we assign objects from our weighted spectroscopic and COSMOS photometric redshift samples and objects targeted by the VVDS Deep survey. Within each four-dimensional k-means cell we find the fraction of the VVDS Deep targets that was successfully assigned a high confidence redshift (flag 3, 4, 9, 13, 14 or 19). In Fig. \ref{spec_incompl} we show the number of VVDS Deep targets and success rate (completeness) in colour-colour space for three ranges in $i$-band magnitude. Between them, these magnitude ranges cover the peak of the number counts in the shear catalogue. 

At relatively bright magnitudes ($i<22.5$) the overall completeness is relatively high, but even here there are typically $20\%$ or more of the targeted galaxies that we do not know the redshifts for. If the incompleteness is due to the clear spectral features of the remaining $20\%$ falling outside of the spectroscopic window then it is easy to imagine that the weighted redshift distribution representing this region of colour-magnitude space would be biased.
At fainter magnitudes the incompleteness increases, first for the reddest objects, but eventually at $i>23$, the majority of subsamples are less than $50\%$ complete.
We cannot remove weak lensing galaxies in all of the incomplete cells without discarding the majority of our sample. Instead, we try to estimate the likely impact of this incompleteness and in particular whether the uncertainties on the inferred means are consistent with the rest of the uncertainties that we estimate in this work. 

In order to estimate the possible impact of incompleteness on the mean redshift of the population we split the colour space cells shown in Fig. \ref{spec_incompl} into regions we term `good' and `bad'. The regions are divided at a completeness of 65\%, which is the median value of the completeness in the cells. We then compare the mean redshift of the weighted spectroscopic sample to the mean from the photometric redshift catalogue published by \cite{ilbert2009} in the COSMOS field, ensuring we use the matched cuts from section \ref{sec:cosmosdata}. Due to the fact that the spectroscopic sample contains many more bright objects than faint, only one quarter of the $\sim40,000$ spectroscopic objects are contained in `bad' cells. We find the difference in the means of the 'good' sample is $\delta z = 0.013$, while $\delta z = 0.03$ for the `bad' regions. These errors are comparable to the expected Poisson errors (which alone should be at the level 0.01) and sample variance (at the level of 0.03), which for a COSMOS sized survey dominates over Poisson errors for samples with more than 1000 galaxies \citep[see Appendix A of ][]{Bordoloi}. For the sample as a whole we therefore do not find evidence for biases in the mean at the level of precision allowed by the samples available. 

Later, in section \ref{sec:wl}, we will see that lensing measurements tend to be dominated by galaxies at higher redshifts. These in turn tend to come from regions with lower levels of completeness. To study this briefly we repeat the comparison between the weighted spectroscopic estimates and the COSMOS samples by first selecting galaxies from the highest redshift bin that we study later ($0.83<z<1.3$, see Sec. \ref{sec:tomo}). We find differences in means of $0.015$ and $0.05$ for the good and bad regions respectively. The samples for this study are significantly smaller. The good regions have 624 and 4255 galaxies in the spectroscopic and COSMOS samples, respectively, and the difference in their means can be explained by Poisson errors alone. The bad regions have 1507 and 17322 galaxies and so the difference in the mean between the spectroscopic and COSMOS determinations cannot be fully explained by Poisson errors alone. However, like the full sample considered above, the difference is similar to that expected from sample variance. We thus conclude that any errors coming from incompleteness for the studies used in this paper are likely to be below the 5\% level.

\subsection{Biases due to template colour coverage}\label{section:templates}

\begin{figure}
\centering
\includegraphics[width=\columnwidth, trim = 0.4cm 1.8cm 1cm 1.8cm, clip]{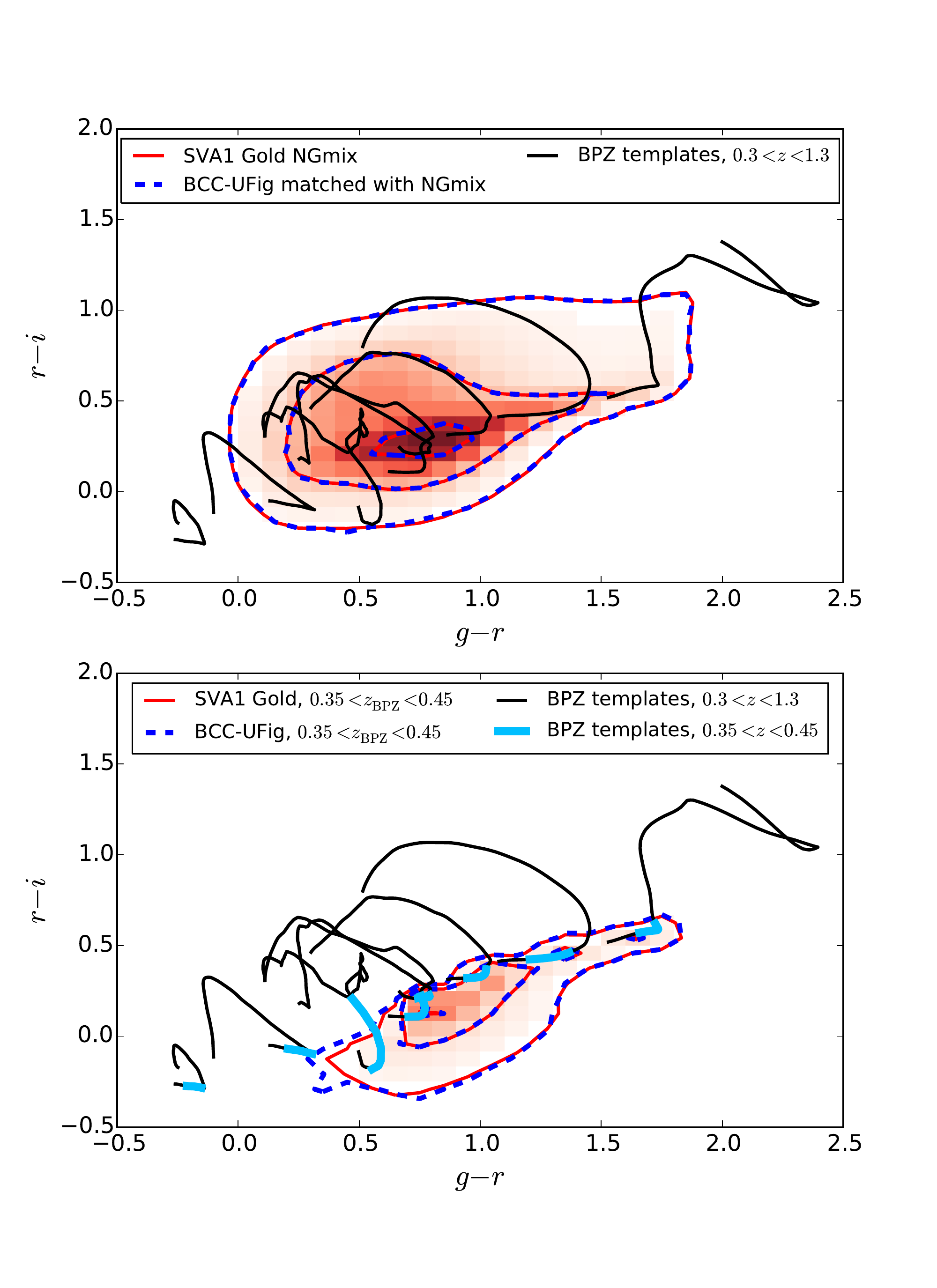}

\caption{
Upper panel: Colour-space distribution of weak lensing sample galaxies and the matched sample taken from BCC-UFig in logarithmic number density intervals (red and blue contours respectively). 
Over-plotted are the observer-frame colours of redshift-evolved galaxy templates (black lines). Here we show the default set of templates included in the \bpz\ photometric redshift code, restricted to  $0.3<z<1.3$ for clarity. Lower panel: The weak lensing and BCC-UFig samples are restricted to objects with \bpz-derived mean redshifts in the range $0.35<z<0.45$. The bold light blue sections of the template tracks indicate the same redshift interval for the galaxy models.
}
\label{template_colour}
\end{figure}

An alternative approach to estimating redshifts empirically based on spectroscopic training samples (\eg via a ML technique) is to use a set of galaxy templates to fit for galaxy redshifts. 
By capturing the rest-frame properties of galaxy spectral energy distributions (SEDs), this modelling approach has the advantage that it can be used to interpolate over regions where there are gaps in spectroscopic samples and to extend to higher redshifts. However, as with all modelling approaches there is a risk of introducing model biases if the templates used for the fitting are not fully representative of true galaxies. 

In this work we have focused on the \bpz template set of \cite{Benitez2000, Coe2006}\footnote{Multiple template-fitting codes and template sets were used in the preparation of this work, though we present a single choice for brevity.} which like many templates, are built for $z=0$ galaxies. 
These do not explicitly account for evolution of the red sequence or changing dust properties at high-z.
The upper panel of Fig. \ref{template_colour} shows the colour space distribution of the weak lensing galaxy sample (red contours) and the matched BCC-UFig sample (blue dashed contours) compared with the observed-frame colours of the \bpz templates redshifted over the range $0.3<z<1.3$. 

Sharp and strong features in galaxy SEDs, such as the $4000{\rm \AA}$ break, create an outer envelope of template colours in certain colour-colour projections. Of particular importance to this work is where the $4000{\rm \AA}$ break transitions between the $g$ and $r$ DECam filters, resulting in extrema in the colours of many templates at $z\sim0.4$. The effect of this is shown in the lower panel of Fig. \ref{template_colour}, where red bold sections of the template tracks correspond to the redshift interval $0.35<z<0.45$. There is clearly a fairly large region of colour-colour space, to the bottom-right of the envelope sampled by the template set, for which the closest template will be at $z\sim0.4$ (in this projection at least). We plot contours for the weak lensing and BCC-UFig samples that lie within this same $0.35<z<0.45$ range, showing that indeed the vast majority of galaxies in this region have a redshift solution at $z\sim0.4$. Previous efforts have in part circumvented this problem, even when using the same template set, by the addition of further photometric bands, in particular $u$-band. Expanding the wavelength coverage with additional bands reduces the reliance on single informative colours for redshift determination. In this way, potential bias introduced from template-fitting is reduced.
For the DES SV data the $u$-band is not observed, but in section \ref{sec:photoz_glob} we show how we use the BCC-UFig simulations to correct to first order for this effect due to the templates colour coverage.

\section{Global Photo-$z$ behaviour and performance}
\label{sec:photoz_glob}

Given the inherent challenges and potential biases in estimating redshifts, we have implemented a number of independent methods for estimating the redshift distribution of the DES SV shear catalogue. Beginning with the global galaxy distribution, we adopt three approaches. The first is an empirical approach based on machine learning methods using spectroscopic training. The second approach is model-based and uses a combination of galaxy templates and calibration using the BCC-UFig simulations. Finally we also estimate the galaxy distribution by matching to COSMOS \photoz data. Agreement between the results can give us confidence that possible systematic errors are subdominant, and the level of discrepancy gives an indication of the level of uncertainty that propagate through to later cosmological constraints. 

\begin{itemize}
\item{\bf Empirical spectroscopic:} Several machine learning \photoz methods have been explored within the DES collaboration, some of which have been previously described in \cite{sanchez2014dessva1photoz}. In the work that follows we focus on a subset of these methods, namely \annz, \skynet\ and \tpz, which are described in more detail in Appendix \ref{app:methods}.
We note that \tpz and \skynet do not  use the weights in training while \annz calculates its own weights that it uses in training.    
\item{\bf Modelling:} For the model-based approach we have implemented the template based method \bpz.
We construct the prior as described in \cite{Benitez2000} by fitting to the training sample of the weighted \msc. Using the same prior presented in \cite{sanchez2014dessva1photoz} has little impact on the results.
To calibrate this method we employ a simple first-order correction by applying weak lensing selection cuts to the BCC-UFig catalogues (see section \ref{sec:bccufig}) and measuring the offset of the mean redshift between these galaxies and that estimated from the pure \bpz $n(z)$. We find this offset to be $0.050$.\footnote{Though the BCC-UFig sample is colour matched to the weak lensing sample after performing the initial weak lensing cuts, this does not influence the correction. If we do not colour-match the BCC-UFig sample to the weak lensing sample, we find an offset of $0.049$.} This offset is applied as a shift to all the \bpz results below, \ie $n(z) \rightarrow n(z-\delta z)$, unless stated otherwise, and is designed to counteract, to leading order, the effect of the peak at $z\sim0.4$ due the template coverage issues (see section \ref{section:templates}) that is present in both the SV data and simulations.
\item{\bf Empirical photometric:} The COSMOS field has been observed using DECam and processed through the DES Data Management pipeline to produce coadd images of similar depth to the main SV survey field. 
Galaxies detected in these images are matched to the \cite{ilbert2009}  \photoz catalogues and then cuts designed to replicate weak lensing selection are applied, as outlined in section \ref{sec:cosmosdata}. Though the \photoz estimates for the COSMOS galaxies are far better than those we can derive from the 5 DES bands, this approach is limited by sample variance.
\end{itemize}
For all the results presented in the sections that follow, we retain $0.3 < {\rm z_{SkyNet}} < 1.3$  galaxies only. 
Redshifts of galaxies outside this range are both poorly estimated and have very little impact on the lensing measurements. Galaxies at low redshift have little lensing signal and there are so few at higher redshift that they can be dropped from the analysis. The redshift cuts are made using the \skynet\ mean, since we have baselined this method as our default, but results that we present are robust to this choice.

\begin{figure}{!htb}
\begin{center}
\includegraphics[width=\columnwidth]{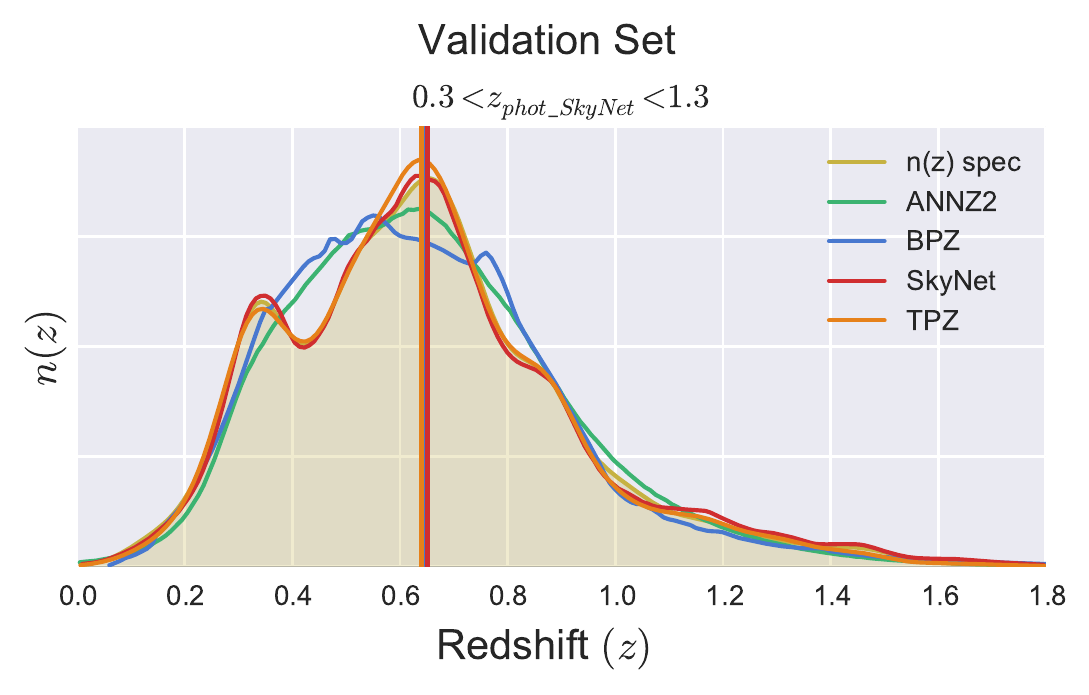}
\includegraphics[width=\columnwidth]{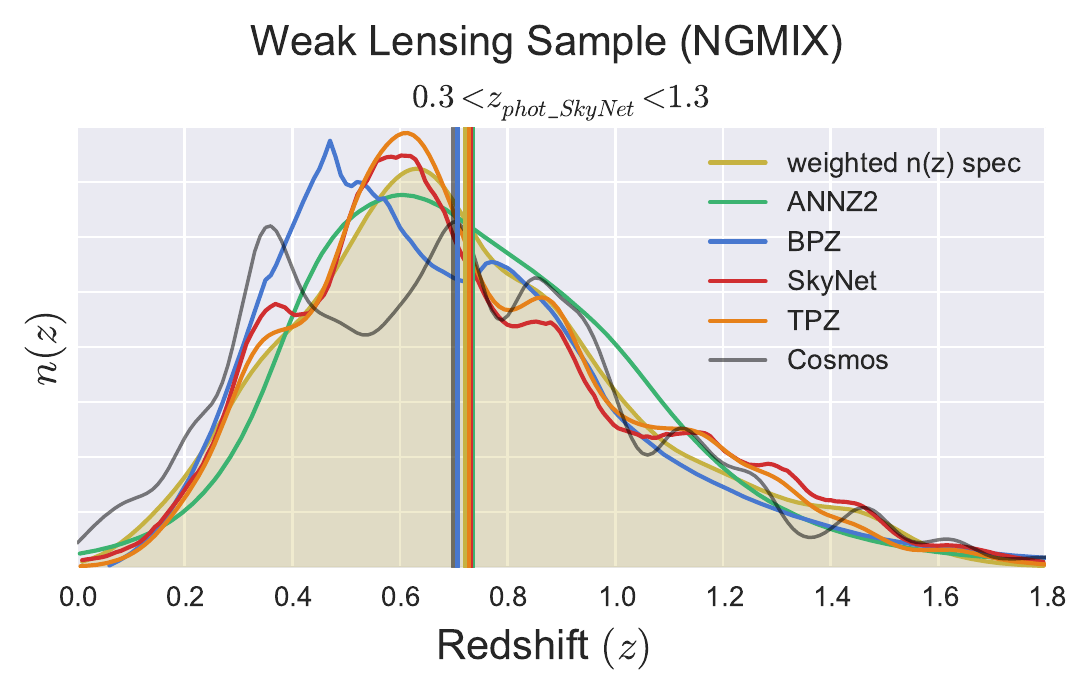}
\end{center}
\caption{The full redshift distribution $n(z)$ for the validation sample ($0.3<z<1.3$). Upper panel: The kernel density estimate of the full unweighted validation sample compared to the four \photoz methods. Lower panel: The same, but including the weighting from Sec. \ref{sec:weights} and matched COSMOS photometric redshifts from \protect\cite{ilbert2009}. 
The vertical lines in the plots are the mean values of the distributions.
}
\label{fig_global}
\end{figure}

The lower panel of Fig. \ref{fig_global} 
shows our reconstruction of the $n(z)$ for the DES SV weak lensing sample. The yellow curve comes from the weighted validation set spectra, which is in effect also an estimator of the global distribution. We also show the results of the three machine learning methods, the modelling based method using \bpz and BCC-UFig and the matched COSMOS results. The vertical lines in the plot show the means of the distributions, which are also listed in Table \ref{table:global}. We focus on the mean since it is well known that uncertainty in the mean is the first order cause of systematic errors in weak lensing \citep{2007MNRAS.381.1018A}. Later, in section \ref{sec:wl}, we will propagate the full errors through to weak lensing statistics and $\sigma_8$. We see that all of our estimates of the global distribution of galaxies give comparable results and we estimate the mean to be $0.72$ with a precision better than 0.02. As a further test, we also show results when we apply the same procedure to the unweighted validation sample. 
Here we take the spectroscopic sample to be a truth catalogue and we can see again that our methods are able to find the mean of this distribution to a precision better than 0.01. The corresponding means for these results are also shown in Table \ref{table:global}. 

\begin{table}
\begin{tabular}{lcc}
\toprule
 & DES SV -WL sample & Validation sample\\
\midrule
Spectra     &  0.72 (weighted)        &   0.64 \\
\annz       &  0.73                   &   0.65 \\
\skynet      &  0.73                  &   0.65\\
\tpz         &  0.73                  &   0.64\\
\midrule
\bpz         &  0.71                  &   0.64 \\
\midrule
Matched COSMOS & 0.70  &   - \\

\bottomrule
\end{tabular}
\caption{
The left column contains the estimates of the mean of redshift distribution of the \ngmix sample of the four photo-z methods and also the
mean of the weighted spectroscopic sample which is itself an estimate of the mean of the \ngmix sample.
The right column contains the mean of the unweighted validation set with the four photo-z methods and the mean from the spectra.
\label{table:global}}
\end{table}

\section{Tomographic Photo-z Performance}\label{sec:tomo}

In the previous section, we discussed the global characteristics of the estimated $n(z)$. 
In the cosmological analysis of \cite{cosmologypaper}, we have presented a conservative analysis of the two-point cosmic shear constraints on cosmology by marginalising over a large array of nuisance parameters related to known or suspected systematics. 
Particularly in the case of intrinsic alignment, doing so severely degrades the constraining power of a non-tomographic analysis. Thus we must also characterise how well the four \photoz methods are able to reconstruct the redshift distribution of individual tomographic bins -- in this case, three bins selected that match those used in \cite{beckeretal2015,cosmologypaper}. 
These are designed to contain approximately equal lensing weight in the larger \ngmix\ shear catalogue.
The bin boundaries are set by cuts on the \skynet mean redshifts at [0.3, 0.55, 0.83, 1.3].
We choose to keep the galaxies in each bin fixed according to the cosmology analysis of DES et al. 2015. 

In this section we look at the \photoz performance in these three tomographic bins. 
This is done through a series of tests, comparing the reconstruction of $n(z)$ (and in particular the value of the mean redshift) in three spectroscopic galaxy samples and the \ngmix\ catalogue:

\begin{itemize}
\item \textbf{Test 1:} An independent sample of spectroscopic galaxies in the VVDS-F14 field, which were not used in training or validation and located in a distinct part of the sky separate from the training and validation fields. The radial structure in the independent sample is thus different from what the machine learning methods trained on.
\item \textbf{Test 2a:} A deeper spectroscopic sample of 30\% of the galaxies in the VVDS-Deep field, which matches better to the depth of DES SV photometry, but which is also part of the validation sample and thus not fully independent.
\item \textbf{Test 2b:} The full validation sample -- 30\% of the matched spectroscopic sample set -- excluding galaxies in the VVDS-F14 field.
\item \textbf{Test 3:} Comparison of the redshift estimates of the four \photoz methods for the full DES SV \ngmix\ catalogue.
\end{itemize}

\begin{figure*}
\centering
\includegraphics[scale=0.8]{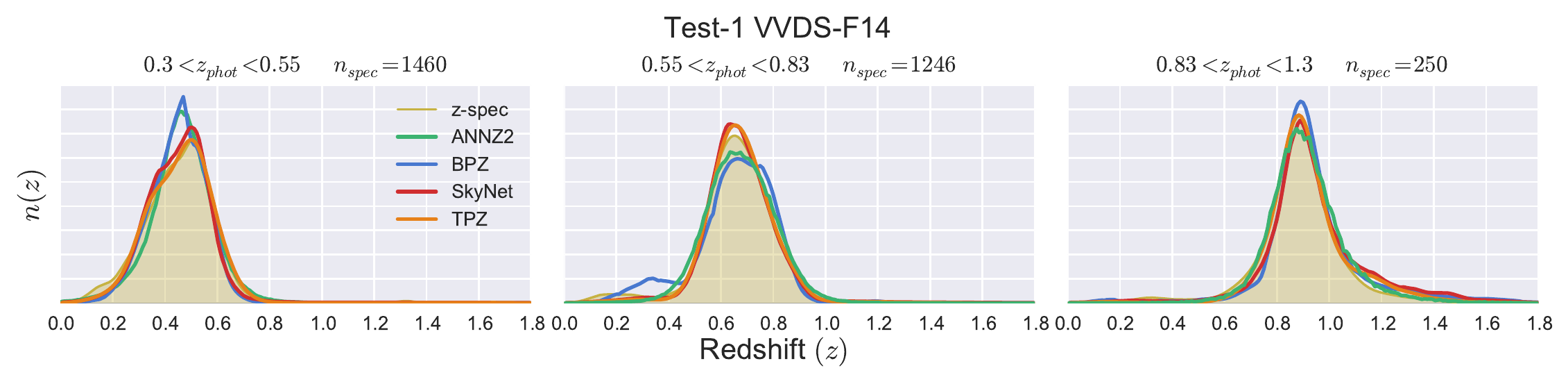}
\includegraphics[scale=0.8]{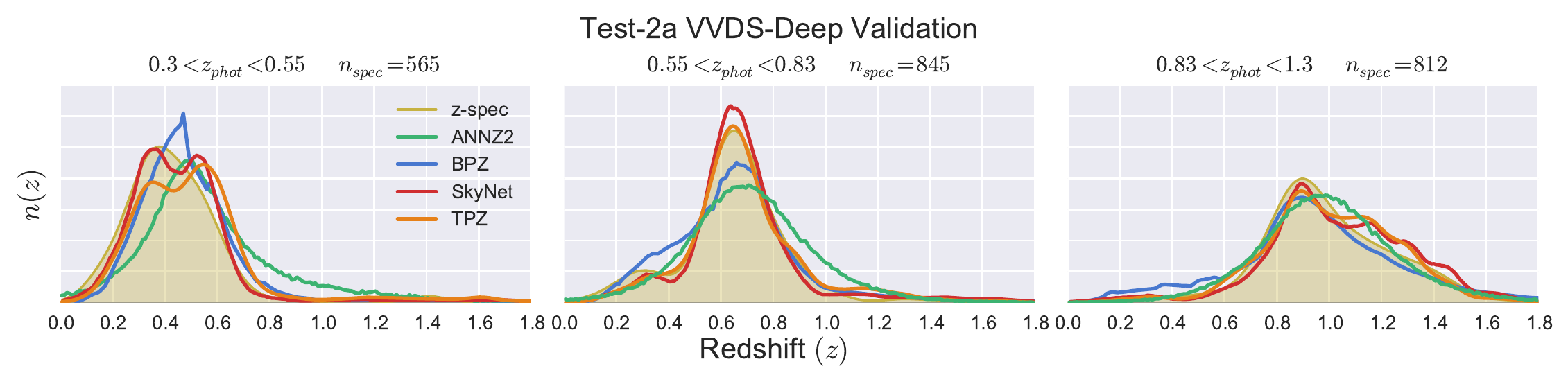}
\includegraphics[scale =0.8]{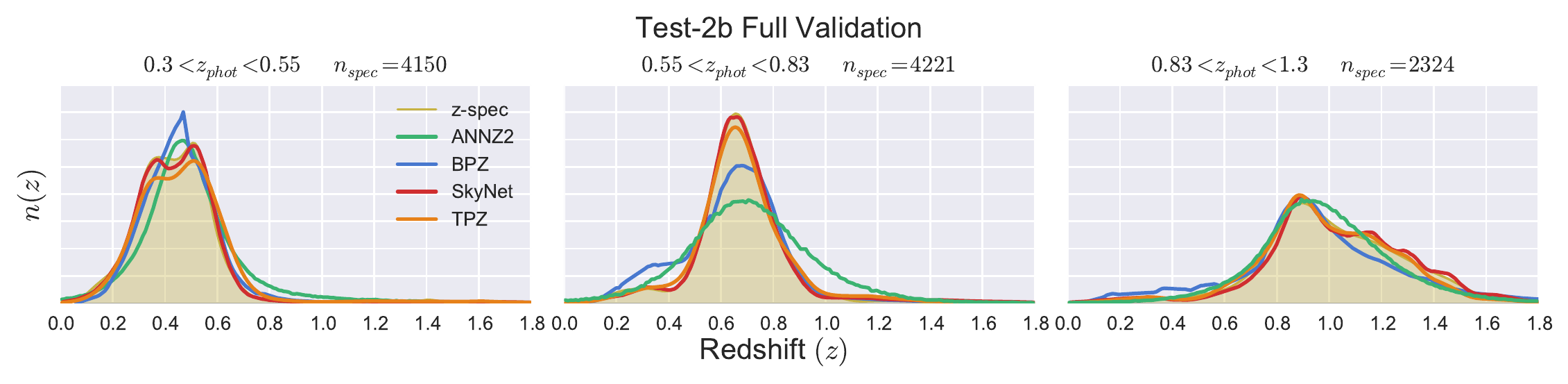}
\includegraphics[scale=0.8]{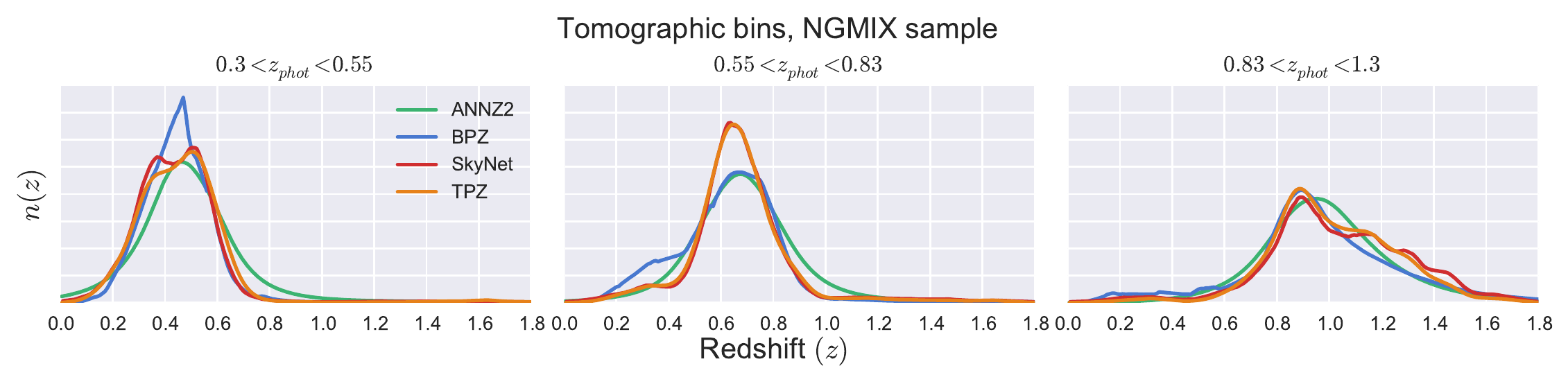}

\caption{Each row of panels show the weighted spectroscopic redshift distributions (shaded area) of the objects in each tomographic bin as selected by the mean of \skynet\ compared to estimates of the redshift distribution of the four methods used in this work. Top row: The spectra used in this test comes from VVDS-F14, an independent sample not not used for training. Second row: The spectra used in this test are a 30\% subset of VVDS-Deep used as part of the validation sample. Third row:  The spectra used in this test are a 30\% subset of the \msc\ used for validation. Bottom row: The redshift distribution in the tomographic bins for the \ngmix\ sample. }
\label{fig:tomo_full}
\end{figure*}

Once again, we use \skynet\ as the fiducial \photoz result, and so for consistency all objects in this section are assigned a bin based on the mean of the \skynet\ $p(z)$. 
In Appendix \ref{sec:app_tomo}, we show results where each code assigns a bin to each galaxy based on their own z-mean. 
Figures \ref{fig:tomo_full} show the results in the tomographic bins of tests 1, 2a and 2b for each of the \photoz algorithms we consider as labelled.
Overall we see that all the methods produce consistent results. 
Since we do not have a perfectly representative spectroscopic sample for the galaxy population for the full \ngmix catalogue, we only compare the relative agreement of the \photoz methods in the bottom panel of Fig. \ref{fig:tomo_full}.
The bin with the highest cosmological information content for tomographic lensing is the highest redshift bin. 
It is therefore reassuring that visually the different methods give consistent results. Table \ref{table:test12a2b} shows the mean offsets of the results shown in the top 3 panels of Fig. \ref{fig:tomo_full}. 
Table \ref{table:test3} shows the estimates of the mean in the tomographic bins of the \ngmix sample by the photo-z codes and the estimate of the weighted spectroscopic sample. 
We see from the results for Tests 2b and 3, which are the closest to our weak lensing samples, that the relative bias of the means are broadly consistent with Gaussian scatter of width 0.05.

\begin{table}
\begin{center}
\begin{tabular}{ccrrrr}
\toprule
& z range& \annz &  \bpz &    \skynet &    \tpz \\
\midrule
\multirow{3}{*}{Test 1} &$0.30 - 0.55 $   & 0.014  &  0.001   &  0.003  &  0.008  \\  
&$0.55 - 0.83 $                           & 0.019  & -0.002   &  0.017  &  0.017  \\ 
&$0.83 - 1.30 $                           & 0.033  &  0.057   &  0.063  &  0.039  \\
\midrule
\multirow{3}{*}{Test 2a} & $0.30 - 0.55 $    &  0.139  &   0.072  &  0.027  &   0.079  \\ 
&$0.55 - 0.83 $                              &  0.069  &   0.027  &  0.034  &   0.042  \\  
&$0.83 - 1.30 $                              &  0.002  &  -0.026  &  0.044  &   0.016  \\  
\midrule
\multirow{3}{*}{Test 2b} &$0.30 - 0.55 $      &  0.064  &  0.032  &  0.012  &   0.033  \\
&$0.55 - 0.83 $                               &  0.027  & -0.010  &  0.013  &   0.010  \\  
&$0.83 - 1.30 $                               & -0.030  & -0.045  &  0.022  &  -0.016  \\ 
\bottomrule
\end{tabular}
\end{center}
\caption{The bias $(\langle z_{phot}\rangle-\langle z_{spec}\rangle)$ between the photometric redshift estimates and the true spectroscopic distribution in Test 1 (`independent'), Test 2a (VVDS-Deep) and Test 2b (Full validation set).}
\label{table:test12a2b}
\end{table}

\begin{table}
\begin{center}
\begin{tabular}{crrrrr}
\toprule
 z range &  Spec                &\annz &  \bpz &    \skynet &    \tpz \\
               &  (weighted)    &           &            &                   &             \\
\midrule
$0.30 - 0.55 $    & 0.45 & 0.49    &  0.46   &  0.45  &  0.46  \\
$0.55 - 0.83 $    & 0.67 & 0.69    &   0.64  &  0.67  &  0.67  \\
$0.83 - 1.30 $    & 1.00 & 0.98    &   0.97  &  1.02  &  1.01  \\
\bottomrule
\end{tabular}
\end{center}
\caption{The estimated mean of the three tomographic bins in the \ngmix sample of the four photo-z methods and the estimate of the weighted spectroscopic sample. }
\label{table:test3}
\end{table}

\section{Implications for weak lensing}
\label{sec:wl}

The mapping of traditional \photoz metrics to actual impacts on the weak lensing measurements and cosmological parameter constraints is non-trivial, and the resulting bias can be difficult to capture using simple metrics. In this section we explore the impact of \photoz uncertainty by propagating the errors through the two-point correlation function and to the cosmological parameter $\sigma_8$ and to measurements of $\Sigma_{\mathrm{crit}}^{-1}$.

\subsection{Photo-z impact on two-point cosmic shear analysis}\label{2pt}

The \photoz $n(z)$ impacts the predicted correlation function (and thus constraints on cosmological parameters) through the lensing efficiency when modelling the convergence power spectrum $C(\ell)$. The tomographic correlation function $\xi_{+/-}$ is related to $C(\ell)$ through the zeroth (fourth) order Bessel function of the first kind by

\begin{equation}
\xi_{+/-,ij}(\theta)=\frac{1}{2\pi}\int d\ell \ell C_{ij}(\ell) J_{0/4}(\ell\theta), \label{eq:xi}
\end{equation}
where $(i,j)\in(1,2,3)$ represent the redshift bins in the auto- or cross-correlation. $C_{ij}(\ell)$ is then defined as

\begin{equation}
C_{ij}(\ell)=\int_{0}^{\chi_H} d\chi \frac{W_i(\chi)W_j(\chi)}{\chi^2}P_{\delta}(\frac{\ell}{\chi},\chi),
\end{equation}
for comoving distance $\chi$, horizon distance $\chi_H$, matter power spectrum $P_{\delta}$, and lensing efficiency, given in a flat universe as
\begin{equation}
W_i(\chi_l)=\frac{3H_0^2\Omega_m}{2c^2} (1+z_l)\chi_l\int_{\chi_l}^{\chi_H}d\chi_s n(\chi_s)\frac{\chi_s-\chi_l}{\chi_s}.
\end{equation}
The redshift distribution of galaxies is normalised such that $\int n_i(\chi) d\chi=1$, $H_0$ is the Hubble parameter, and $\Omega_m$ is the matter density parameter at $z=0$. 

\begin{figure}
\begin{center}
\includegraphics[width=\columnwidth]{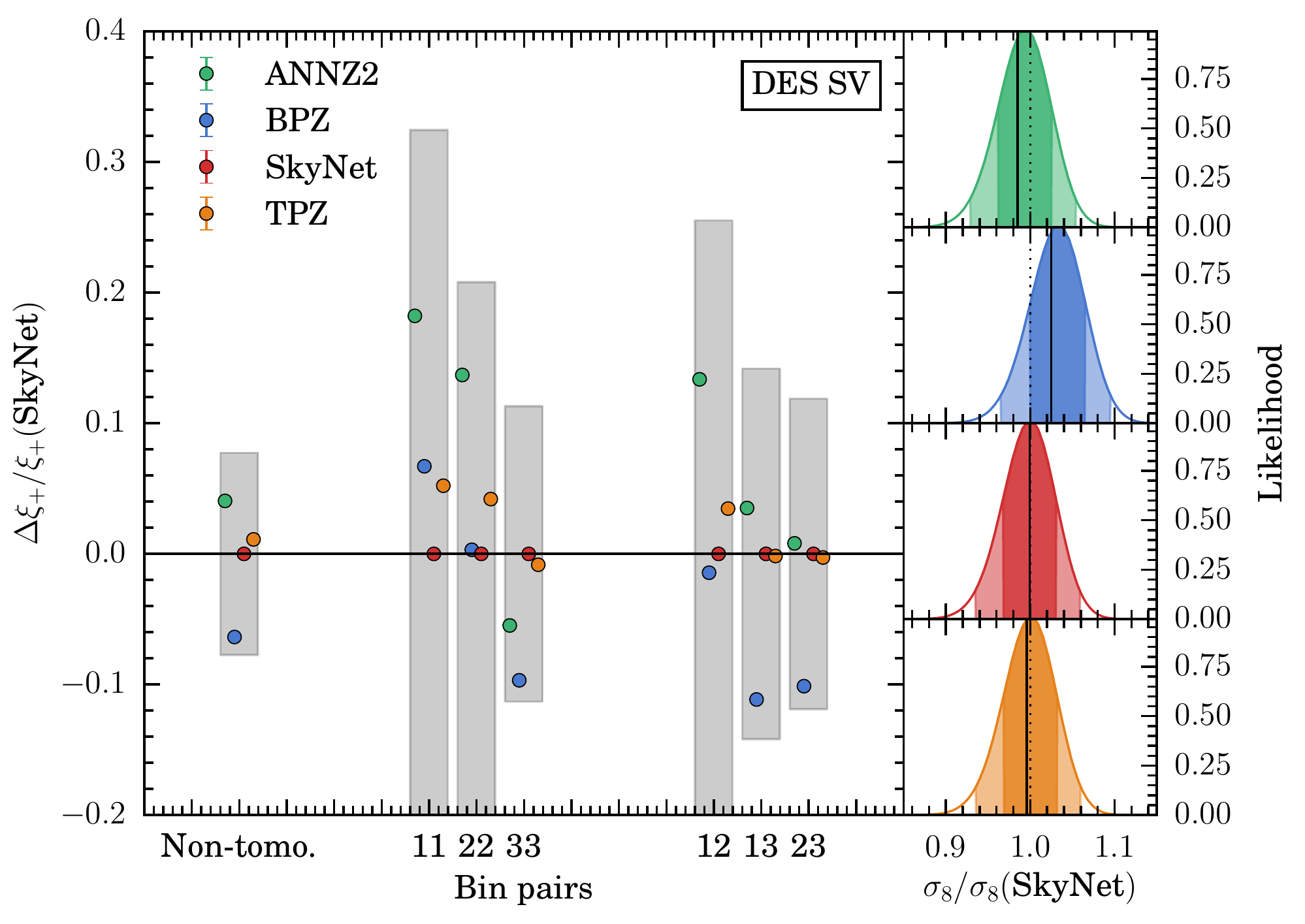}
\end{center}
\caption[]{A comparison of the relative agreement of the $n(z)$ estimates for \annz, \bpz, \skynet, and \tpz~for the \ngmix\ shear catalogue. Left panel: The relative magnitude of the correlation function compared to the spectroscopic $n(z)$ prediction is shown for the non-tomographic $\xi_+$, the three auto-correlations, and the three cross-correlations. The grey band is the actual variance in the magnitude of $\xi_+$ measured from SV data. Right panels: The corresponding constraints on $\sigma_8$, with fiducial \skynet results normalised to one (vertical dotted black line). The likelihood histograms, colour-coded to match the $\xi_+$ points on the left, are shown for each tomographic constraint. The peak of the likelihood histogram for the non-tomographic constraint is given by the vertical black line for comparison. The vertical ordering is the same as the legend in the left panel. \label{fig:xi_full}
}
\end{figure}

The predicted $\xi_{+/-}$ (both tomographic and non-tomographic) are calculated over the $\theta$ range and tomographic binning used for the measurements in \cite{beckeretal2015} for each \photoz estimate and the weighted matched spectroscopic sample. We then use these predicted correlation functions with the covariance matrix from \cite{beckeretal2015} to propagate the differences between photo-z estimates through to constraints on $\sigma_8$ (with all other parameters fixed). The `truth' (or measurement of $\xi_{+/-}$ with no systematic uncertainties) is taken to be either the fiducial \skynet\ prediction in Sec. \ref{wlsv} or the weighted matched spectroscopic sample in Sec. \ref{wlnull}, while each photo-z estimate's predicted $\xi_{+/-}$ is taken to be the assumed theory in turn when constraining $\sigma_8$. The final results of this comparison for the four \photoz estimates presented in this work are shown in Figs. \ref{fig:xi_full} -- \ref{fig:xi_deep}.

\subsubsection{Comparison of \photoz estimates for the DES SV shear catalogue}\label{wlsv}

For the full photometric galaxy sample contained within the shear catalogue, we have no estimate for the true value of the $n(z)$ to compare to and so instead compare to the fiducial \skynet\ prediction as a relative point of reference. 
We can therefore only compare the relative agreement between the \photoz codes shown for the \ngmix\ catalogue in Fig. \ref{fig:xi_full}. 

In the left panel, the relative agreement in the magnitude of $\xi_+$ is shown, averaged over $\theta$.\footnote{The major results are unchanged when instead considering specific values of $\theta$.} The left set of points show the non-tomographic $\xi_+$, while the middle and right sets of points show the three auto- and cross-correlations, respectively. The grey bands show the 1$\sigma$ error on the magnitude of the measured $\xi_+$ for each correlation function, using the covariance calculated in \cite{beckeretal2015}. The relative agreement in $\xi_+$ between the machine learning methods is very good in correlations with the highest tomographic bin (`33', `23', and `13'). This increases significantly for correlations with the lower tomographic bins (`11', `22, and `12'), though the non-tomographic case also has good agreement on the order of 5\%. \bpz tends to disagree with the machine learning methods, typically at the 5-10\% level.

The right panels of Fig. \ref{fig:xi_full} show the corresponding constraints on $\sigma_8$. The \skynet prediction is normalised to one (vertical dotted black line). The likelihood histogram, coloured to match the points in the left panel for each \photoz code, is shown for the full tomographic constraint, while the vertical solid black line gives the peak of the likelihood histogram for the non-tomographic constraint. The bias in constraints on $\sigma_8$ between the machine learning \photoz methods is very small despite low-z differences in the correlation function, with agreement at much better than the 1$\sigma$ level. \bpz has a relative bias of about 1$\sigma$, by comparison, which corresponds to about 3\% in $\sigma_8$.

For completeness, we have also repeated the above analyses and those in Sec. \ref{wlnull} on the \imshape\ $n(z)$ with the same redshift boundaries matching those derived for \ngmix\ and again for tomographic bins derived for \imshape, and find in all cases that the major conclusions and resulting differences across \photoz methods are consistent between analyses of the two catalogues at the level of accuracy we require for SV analysis.

\subsubsection{Null tests relative to matched spectroscopic samples}\label{wlnull}

One difficulty with the results in Sec. \ref{wlsv} is that we have no way of determining what the true $n(z)$ is, and thus can only compare relative agreement between photo-z methods.  We can, however, create an experiment in which the $n(z)$ is known to be exactly that of our weighted independent spectroscopic sample (Test 1). We then repeat the analysis from Sec. \ref{wlsv} for this test as an additional way of characterising systematic photo-z uncertainties. Though there are only 2956 galaxies in the independent spectroscopic sample within our $0.3<z<1.3$ boundaries, we assume the estimated $n(z)$ from each code and the spec-z distribution instead represents a sample with the same number of objects as the \ngmix~catalogue. These redshift distributions (see top panel Fig. \ref{fig:tomo_full}) are used to measure the relative difference in $\xi_{+/-}$ compared to the spectroscopic prediction as in Sec. \ref{wlsv}. We also calculate error bars on the points, which represent the $1\sigma$ error in the difference from bootstrapping the $n(z)$ of the sample. Since we are comparing the matched photometric and spectroscopic $n(z)$ distributions for the same galaxies contained within the VVDS-F14 field, there is no sample variance contribution to these error bars. However, since it is a small field separate from the DES SV SPT-E region, any extrapolation of the bias to the full DES SV shear catalogue could still be over- or under-estimated. 

\begin{figure}
\begin{center}
\includegraphics[width=\columnwidth]{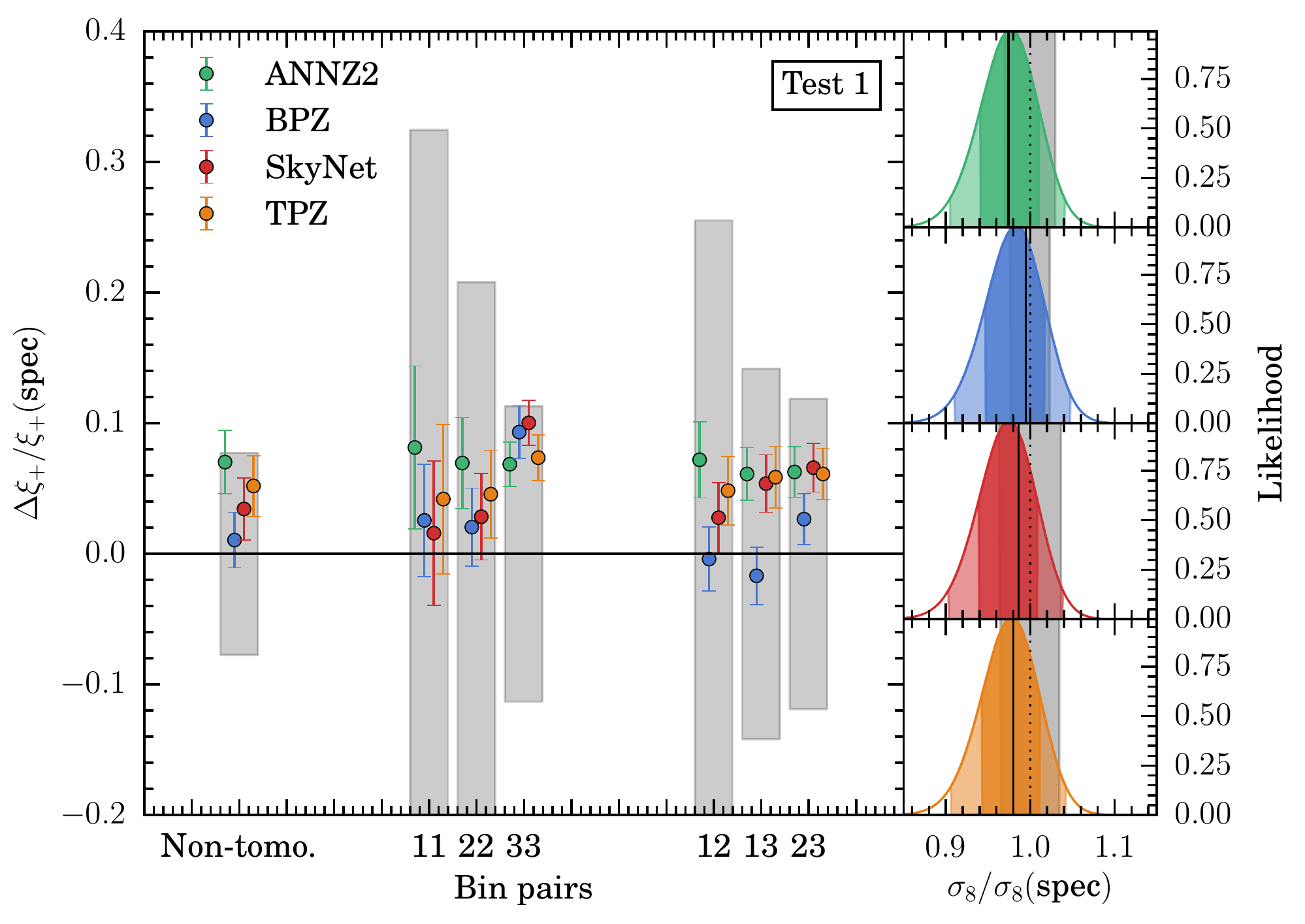}
\end{center}
\caption[]{A comparison of the relative agreement of the $n(z)$ estimates for \annz, \bpz, \skynet, and \tpz~to the weighted independent spectroscopic galaxy sample. Left panel: The relative magnitude of the correlation function compared to the spectroscopic $n(z)$ prediction is shown for the non-tomographic $\xi_+$, the three auto-correlations (11, 12, 33 bin pairs), and the three cross-correlations (12, 13, 23 bin pairs). The grey band is the actual variance in the magnitude of $\xi_+$ measured from SV data. Error bars on the points are the 1-$\sigma$ error on the difference of $\xi_+$ obtained from bootstrapping the $n(z)$ of the spectroscopic sample. Right panels: The corresponding constraints on $\sigma_8$, normalised to one (vertical dotted black line). The likelihood histograms, colour-coded to match the $\xi_+$ points on the left, are shown for each tomographic constraint. The peak of the likelihood histogram for the non-tomographic constraint is given by the vertical black line for comparison. The vertical grey band is the corresponding 1-$\sigma$ bootstrap error in $\sigma_8$. \label{fig:xi_blind}}
\end{figure}
\begin{figure}
\begin{center}
\includegraphics[width=\columnwidth]{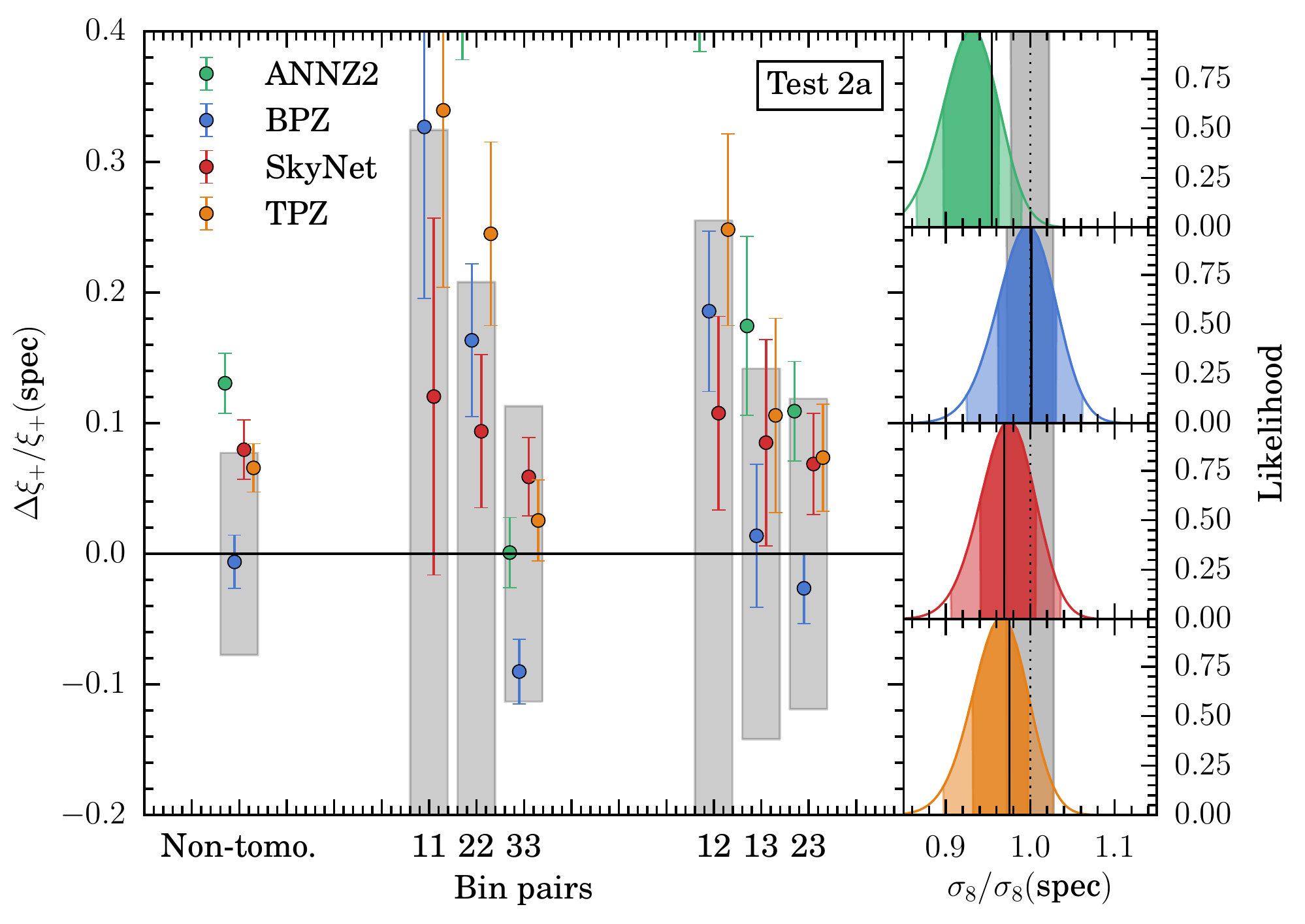}
\end{center}
\caption[]{A comparison of the relative agreement of the $n(z)$ estimates for \annz, \bpz, \skynet, and \tpz~to the weighted 'deep' spectroscopic galaxy sample, showing the same information as described in Fig. \ref{fig:xi_blind}.  \label{fig:xi_deep}}
\end{figure}

\begin{table*}
\begin{tabular}{c|c|cccc}
\toprule
 && \annz &  \bpz &    \skynet &    \tpz \\
\midrule
\multirow{4}{*}{Test 1}& \annz &  -- &  0.36 (-0.04) &  0.22 (-0.02) &  0.03 (-0.01)  \\
& \bpz &  -0.36 (0.04) &  -- &  -0.13 (0.01) &  -0.33 (0.03)  \\
 &\skynet &  -0.22 (0.02) &  0.13 (-0.01) &  -- &  -0.2 (0.01)\\
 &\tpz &  -0.03 (0.01) &  0.33 (-0.03) &  0.2 (-0.01) &  -- \\
\midrule
 \multirow{4}{*}{Test 2a}&\annz &  -- &  -3.94 (-0.1) &  -7.02 (-0.04) &  -5.2 (-0.06)   \\  
 &\bpz &  3.94 (0.1) &  -- &  -3.08 (0.07) &  -1.26 (0.04)   \\ 
 &\skynet &  7.02 (0.04) &  3.08 (-0.07) &  -- &  1.82 (-0.02)    \\ 
 &\tpz &  5.2 (0.06) &  1.26 (-0.04) &  -1.82 (0.02) &  -- \\
\midrule
&\annz &  -- &  -0.08 (-0.02) &  -0.08 (-0.02) &  -0.04 (-0.01) \\ 
 Test 2a &\bpz &  0.08 (0.02) &  -- &  0.0 (-0.0) &  0.04 (0.01)  \\ 
 Corrected&\skynet &  0.08 (0.02) &  -0.0 (0.0) &  -- &  0.04 (0.01) \\ 
 &\tpz &  0.04 (0.01) &  -0.04 (-0.01) &  -0.04 (-0.01) &  --  \\  
\bottomrule

\end{tabular}
\caption{Values of $\ln K$ for the Bayes factor $K=Pr(D|p_1)/Pr(D|p_2)$ are shown for each \photoz estimate ($p_1$ - rows) compared to another ($p_2$ - columns) when constraining the value of $\sigma_8$ (all other cosmology is kept fixed, varying only the estimates of $n(z)$ between $p_1$, $p_2$, and $D$). The values for tomographic (non-tomographic) analyses in Figs. \ref{fig:xi_blind}, \ref{fig:xi_deep}, and the right panel of \ref{fig:biastest} are given. The Bayes factor gives an indication of how much more supported one \photoz estimate ($p_1$) is than another ($p_2$) by the data $D$, in this case the predicted correlation function built from the weighted spectroscopic estimate of $n(z)$. A value $\ln K>1$ generally indicates that $p_1$ is more strongly supported as the true \photoz estimate. \label{table:bayes1}}
\end{table*}

We show the results of this analysis in Fig. \ref{fig:xi_blind}. The bias in $\xi_+$ relative to the spectroscopic prediction for the three machine learning codes (\annz, \skynet, and \tpz) is shown in the left panel. It is in good agreement and consistent across the correlations at about $5-10\%$ larger than the spectroscopic prediction. This is consistent with the machine learning codes producing too wide $p(z)$ or over-estimated high-z tails, both of which can bias $\xi_+$ high. The empirically corrected \bpz photo-z estimates perform similarly, with a maximum bias in $\xi_+$ of 10\% in the highest redshift auto-correlation. 

The right panels of Fig. \ref{fig:xi_blind} show the corresponding constraints on $\sigma_8$. The weighted spectroscopic prediction is normalised to one (vertical dotted black line) and the vertical grey band is the $1\sigma$ bootstrap error corresponding to the error bars on the $\xi_+$ points. Note, however, that discussion of deviations in $\sigma_8$ will refer primarily to the marginalised constraints unless specifically referring to the bootstrap error. The tomographic and non-tomographic constraints agree well. All four photo-z estimates are biased slightly low by just less than 1$\sigma$. It is important to note that due to the small sample size in the independent spectroscopic test sample, the $1\sigma$ bootstrap error in $\sigma_8$ just due to sample variance in the independent spectroscopic sample is of the same order as the $1\sigma$ constraints on $\sigma_8$ in DES SV for some methods. Overall, we find a level of systematic bias from this test in $\sigma_8$ of 1-3\%.

We can further diagnose the performance of the \photoz codes' estimates of the $n(z)$ by considering the Bayes factor
\begin{equation}
K=\frac{Pr(D|p_1)}{Pr(D|p_2)},
\end{equation}
where $Pr$ is the posterior probability of the model $p_i$ due to some \photoz estimate in the $\sigma_8$ constraints of Fig. \ref{fig:xi_blind}. In this analysis, $D$ refers to the predicted $\xi_{+/-}$ for the weighted matched spectroscopic samples, and $Pr$ is the integrated posterior likelihood. The Bayes factor can be used to compare how well supported by the data two models are. A value $\ln K>1$ supports $p_1$ over $p_2$, with $p_1$ being substantially supported when $\ln K>3$. The Bayes factor is given for each combination of \photoz estimates in Table \ref{table:bayes1}. The Bayes factors from the tomographic analysis are given first, with the non-tomographic Bayes factors shown in parentheses for comparison. We find that there is no significant preference for one \photoz code over another for the independent sample (Test 1),  though there is some evidence that \annz\ does slightly worse and the corrected \bpz\ slightly better. This distinction is lost, however, for the non-tomographic analysis, which is unable to differentiate the \photoz estimates. 

\begin{figure*}
\begin{center}
\includegraphics[width=\columnwidth]{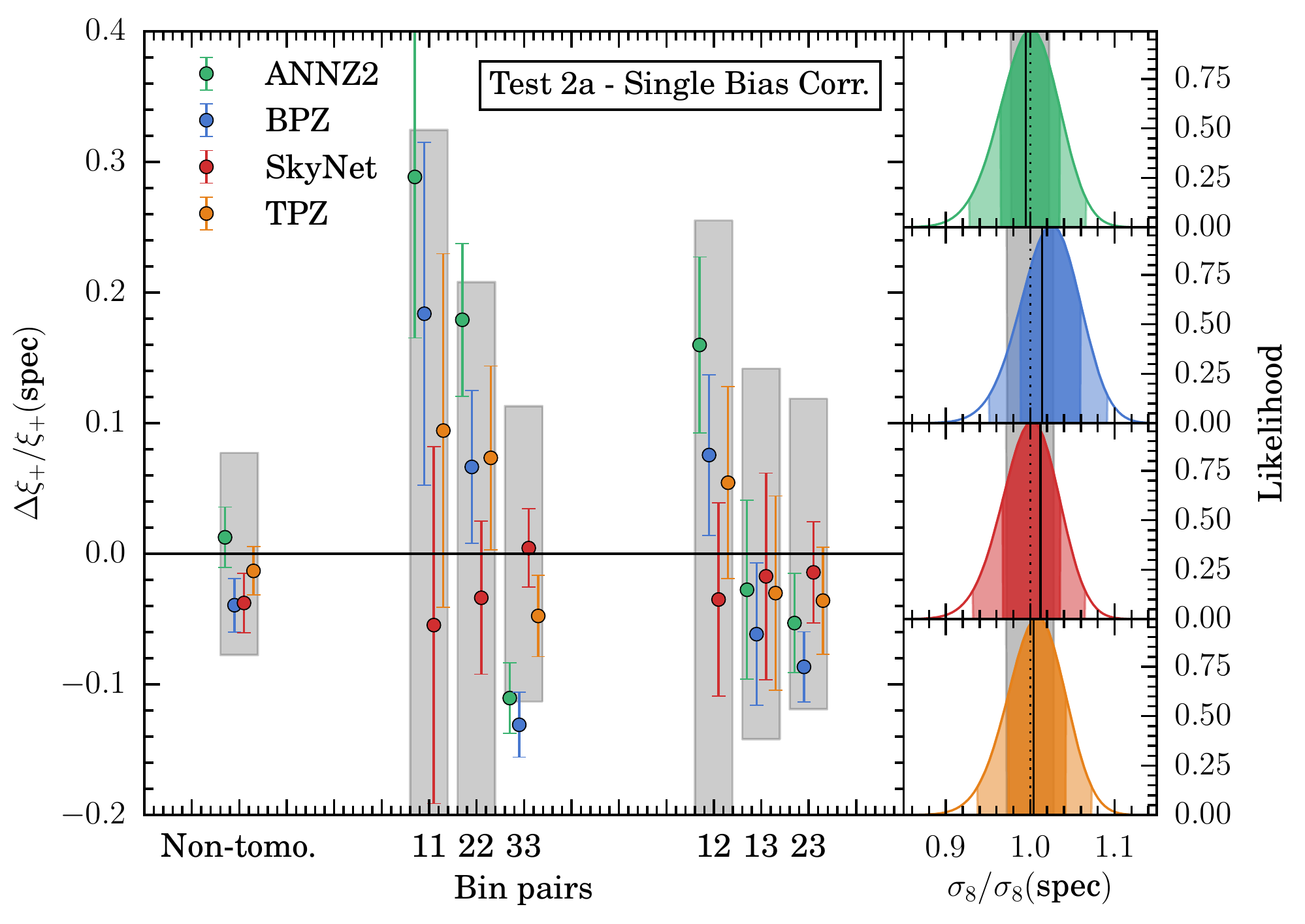}
\includegraphics[width=\columnwidth]{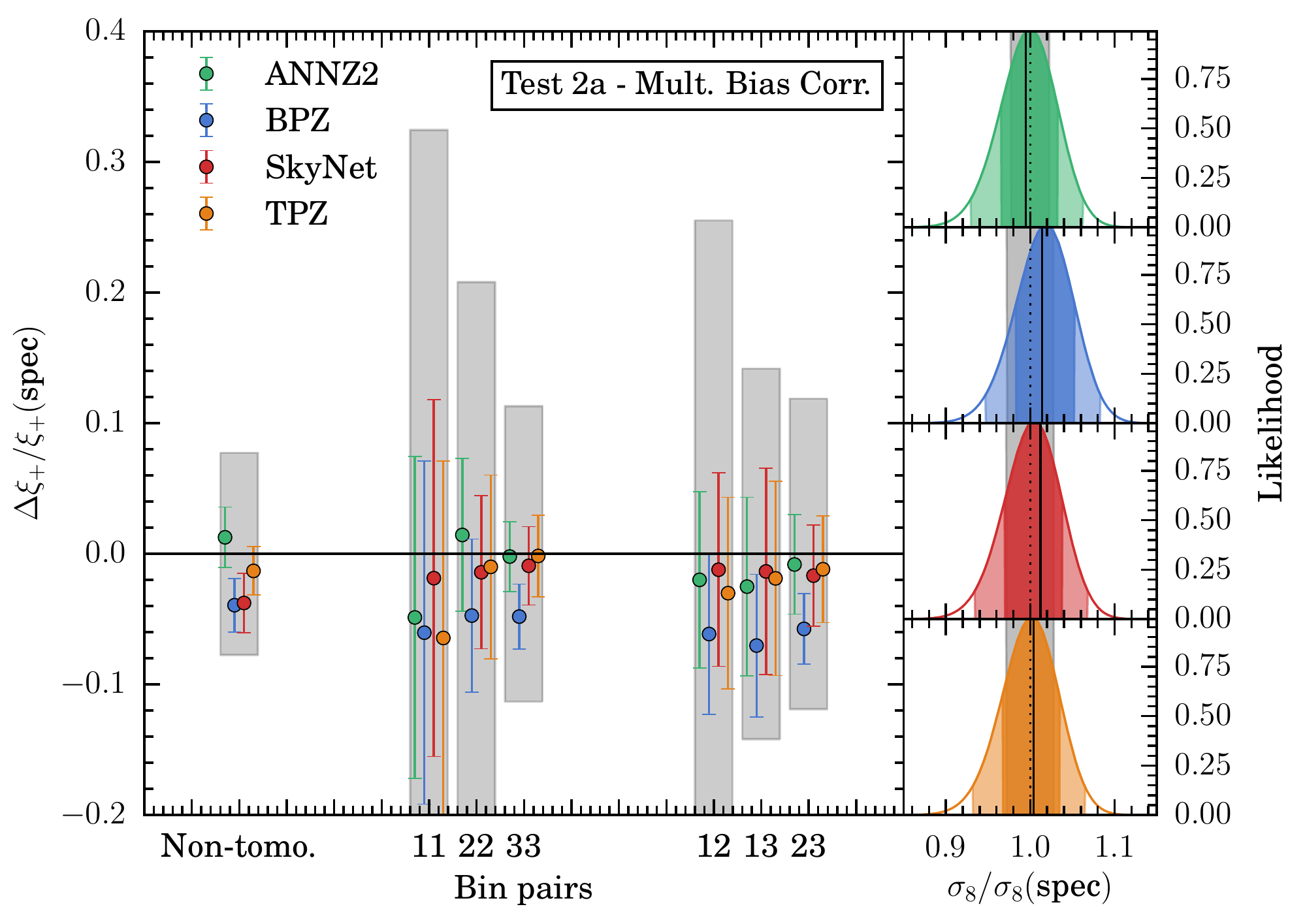}
\end{center}
\caption[]{The effect on Fig. \ref{fig:xi_deep} of applying a bias correction to the mean of the $n(z)$ of each \photoz estimate by comparison to the true spectroscopic $n(z)$. The left side fixes a single bias parameter for the three tomographic bins, while the right side allows a different bias parameter for each bin. Each side shows the same information as described in Fig. \ref{fig:xi_blind}.  \label{fig:biastest}}
\end{figure*}

\begin{figure}
\begin{center}
\includegraphics[width=\columnwidth]{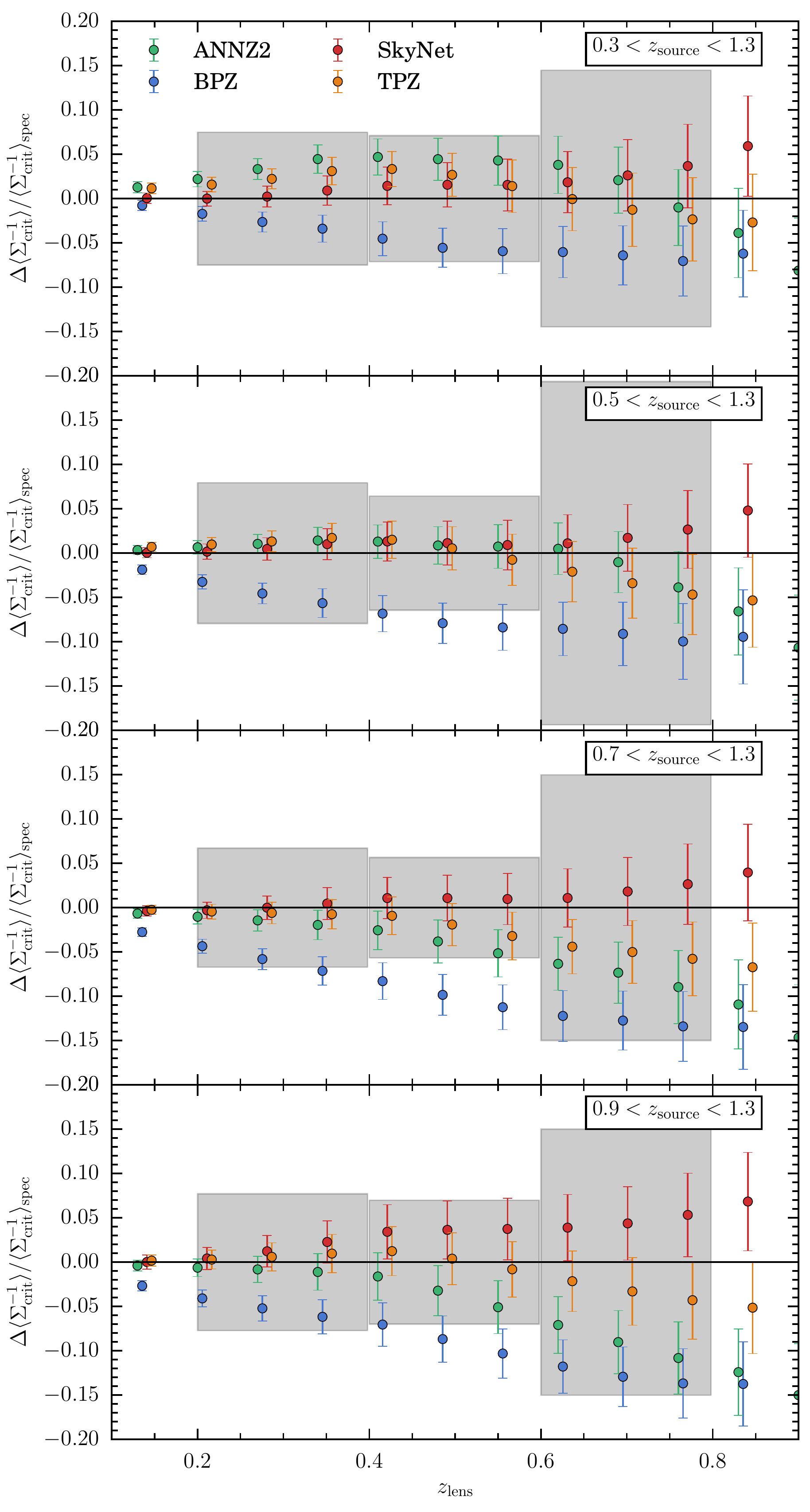}
\end{center}
\caption[]{The fractional difference in the $\langle \Sigma_{\mathrm{crit}}^{-1}\rangle$ between the \photoz estimates and the deep Test 2a spectroscopic prediction is shown as a function of lens redshift for four source redshift bins. Grey bands show the 1$\sigma$ statistical error in the measurement of the tangential shear signal for the three lens bins indicated by the width of the bands. \label{fig:sigma_spec}}
\end{figure}

We also want to compare the \photoz performance of the four codes for a set of spectroscopic redshifts that better match the depth of the DES SV data. Figure \ref{fig:xi_deep} instead compares the correlation function and $\sigma_8$ constraints for the \photoz estimates of galaxies in the weighted 'deep' spectroscopic sample of Test 2a. The predicted $n(z)$ for these galaxies is shown in the third panel in  Fig. \ref{fig:tomo_full}. All four codes perform more poorly for this 'deep' sample compared to the analysis of Test 1 in Fig. \ref{fig:xi_blind}, with a greater spread in the magnitude of the predicted $\xi_+$ relative to the spectroscopic prediction. \skynet\ is the most stable across tomographic bins, with a spread in bias values limited to around 5\%. The other codes scatter to a much wider range of values. For the lower bins in particular, there is significant bias in $\xi_+$.

The corresponding $\sigma_8$ constraints are driven by information in the highest redshift bin, however, and have a more reasonable bias about the weighted spectroscopic prediction. The four \photoz estimates still agree with the matched spectroscopic prediction for the VVDS-Deep sample within 1-$\sigma$, except for \annz, which is biased at the 2-3$\sigma$ level. The large range of bias between the lowest and highest redshift bins also produces a nearly 1$\sigma$ tension between the tomographic and non-tomographic constraints for \annz. This bias is not explained as an artefact of selecting the binning of galaxies based on \skynet, as seen in App. \ref{sec:app_tomo}. We present the associated Bayes factor values in Table \ref{table:bayes1}, where \skynet\ is significantly favoured over the other three codes. There is again no distinction between the codes, however, in the non-tomographic analysis from the Bayes factor. Overall, we find a maximum level of systematic bias from this test in $\sigma_8$ of 7\% for \annz, though the bias in the other methods is similar to the level found in Test 1.

\subsection{Validation of priors for \photoz bias parameters}\label{wlbias}

To first order, we can correct for the systematic redshift biases shown in section 6.1 with the approximation $n_i(z)\rightarrow n_i(z-\delta z_i)$ where $\delta z_i$ is
the bias on the mean redshift of the source galaxies in the appropriate tomographic bin. In the cosmology analysis of DES et al 2015. we adopt a Gaussian prior of width 0.05 on the allowed bias values based on comparisons of the four \photoz method's estimates of the $n(z)$ discussed in Secs. \ref{sec:photoz_glob} \& \ref{sec:tomo}. This is shown explicitly in Fig. \ref{fig:biastest}, where we compare the impact such a correction scheme has on $\xi_+$ and $\sigma_8$. The bias parameters by which the $n(z)$ are shifted are not marginalised over here, but instead are taken from Table \ref{table:test12a2b} for Test 2a, since we can directly calculate the bias.

We find that a single mean redshift bias parameter is sufficient to resolve the bias in $\sigma_8$ for all four codes. Taking into consideration the $1\sigma$ bootstrap error in the $\xi_+$ ratio, all the tomographic correlations are consistent with zero remaining bias in $\xi_+$ for \skynet, and the other \photoz estimates are also greatly improved relative to the spectroscopic prediction. Relaxing this to a bias parameter for each redshift bin does not further significantly improve the bias in $\sigma_8$, but it does have a large impact on the agreement in $\xi_+$, which could have an impact on other parameter constraints. All tomographic points are now consistent with zero for the machine learning methods. This is confirmed in the Bayes factor, shown in Table \ref{table:bayes1} for the three-parameter case. All values of $K$ are consistent with the four corrected \photoz estimates being equally likely to be true. 

We thus employ a Gaussian prior on the photo-z bias of width $\delta z_i=0.05$, centred at zero, separately for each of the tomographic bins in the fiducial cosmology analysis of DES et al. 2015. We also explore the effect of propagating a non-zero centre for the prior in the analysis discussed in that paper, and find no significant differences to the cosmology results. 

\begin{figure*}
\begin{center}
\includegraphics[width=1.75\columnwidth]{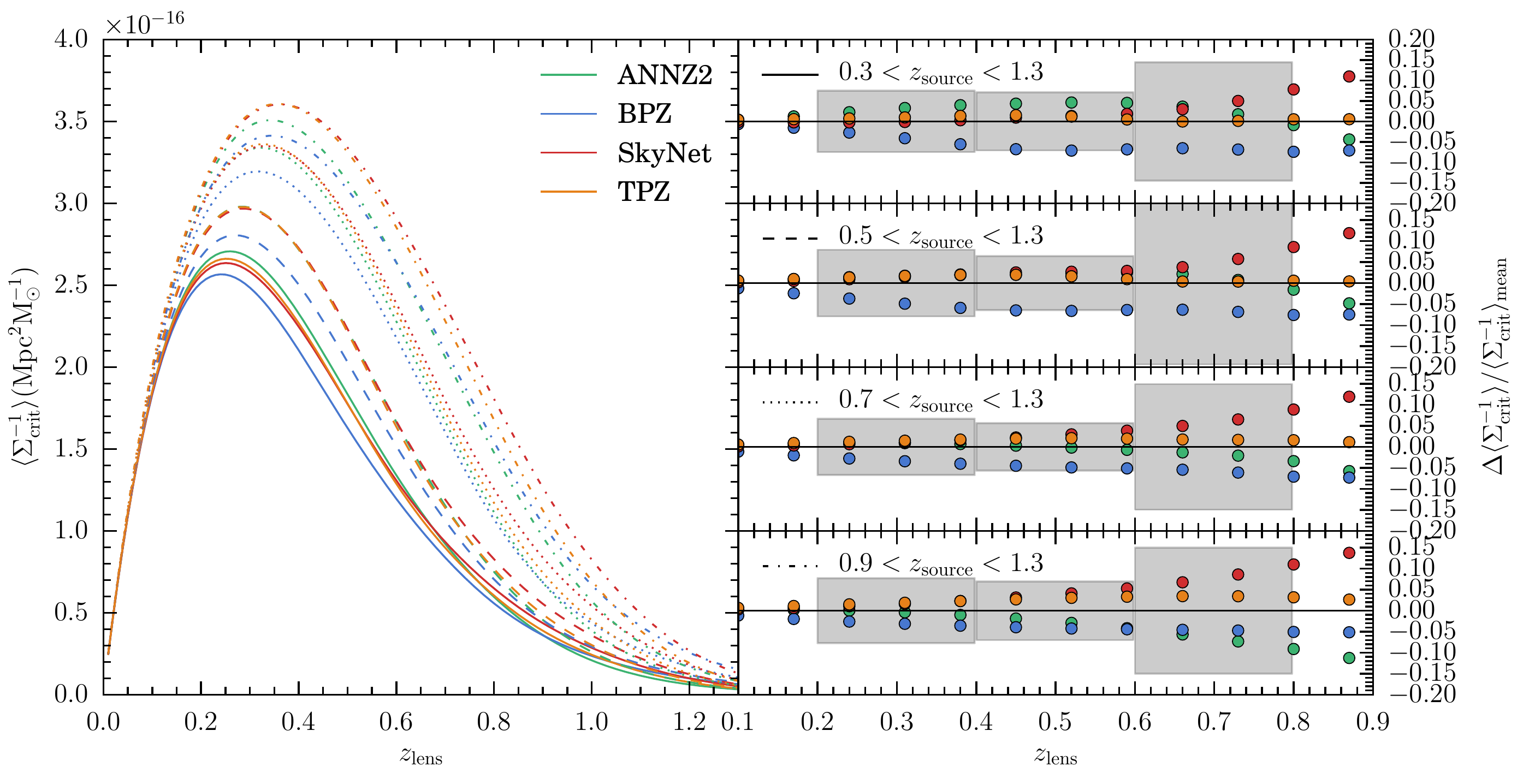}
\end{center}
\caption[]{Left: $\langle \Sigma_{\mathrm{crit}}^{-1}\rangle$ for the full \ngmix shear catalogue is shown as a function of lens redshift for four source redshift bins: $0.3<z_{\mathrm{source}}<1.3$ (solid), $0.5<z_{\mathrm{source}}<1.3$ (dashed), $0.7<z_{\mathrm{source}}<1.3$ (dotted), and $0.9<z_{\mathrm{source}}<1.3$ (dash-dotted). Right: The fractional difference in the $\langle \Sigma_{\mathrm{crit}}^{-1}\rangle$ between the \photoz estimates relative to the mean. Grey bands show the 1$\sigma$ statistical error in the measurement of the tangential shear signal for the three lens bins indicated by the width of the bands. \label{fig:sigma_sv}}
\end{figure*}

\subsection{Photo-z impact on other lensing analyses}\label{tan}

In general, the main impact of photo-z uncertainties in weak lensing measurements enters through the impact on the critical surface density $\Sigma_{\mathrm{crit}}$. This quantity captures the information on distance ratios in lens-source pairs that lensing is sensitive to, namely
\begin{equation}
\Sigma_{\mathrm{crit}}^{-1}=\frac{4\pi G}{c^2}\frac{D_{ls}D_l}{D_s},
\end{equation}
where $D_l$ is the angular diameter distance to the lens, $D_s$ is the distance to the source, and $D_{ls}$ is the distance between lens and source. Calculating $\Sigma_{\mathrm{crit}}^{-1}$ uses the individual $p(z)$ for each galaxy, which is a different test of the \photoz quality than the bulk summation into large tomographic bins for cosmic shear analysis. It is also possible to directly calculate this quantity for a relatively small sample of galaxies, unlike the correlation function, allowing us to directly compare the \photoz methods' predictions for this quantity with the weighted matched spectroscopic prediction. 

To explore this, we compare the impact of the different redshift estimates on the calculation of $\langle \Sigma_{\mathrm{crit}}^{-1}\rangle$ as a function of lens redshift. This directly probes the impact of \photoz bias in measurements of $\Delta \Sigma$ in cluster and galaxy-galaxy lensing, and is relevant for other tangential shear measurements where one can distinguish between some a population with significantly better \photoz estimates than a source sample in the larger shear catalogue. We will assume that lens galaxies have negligible redshift error relative to the source catalogue and thus have no impact on the calculation of $\langle \Sigma_{\mathrm{crit}}^{-1}\rangle (z_{\mathrm{lens}})$ for the purpose of evaluating the redshift estimates presented in this paper. We follow the same process as described above, repeating this analysis for the deep matched spectroscopic sample (Test 2a) and for the full DES SV \ngmix shear catalogue. 

For each galaxy sample and \photoz estimate, we evaluate the weighted mean inverse $\Sigma_{crit}$ as a function of lens redshift

\begin{equation}
\langle \Sigma_{\mathrm{crit}}^{-1}\rangle (z_{\mathrm{lens}})=\sum_i p(z_{\mathrm{source},i})\Sigma_{crit}^{-1}(z_{\mathrm{lens}},z_{(\mathrm{source},i)}),
\end{equation}
where $\sum p(z)=1$. For the spectroscopic test, $\langle \Sigma_{\mathrm{crit}}^{-1}\rangle_{\mathrm{spec}}$ is simply evaluated at the spectroscopic redshift with no probability distribution. We use three source redshift bins: $0.5<z_{\mathrm{source}}<1.3$, $0.7<z_{\mathrm{source}}<1.3$, and $0.9<z_{\mathrm{source}}<1.3$, as well as the non-tomographic range from the two-point analysis, $0.3<z_{\mathrm{source}}<1.3$. We calculate $\langle \Sigma_{\mathrm{crit}}^{-1}\rangle$ over the lens redshift range $0.1<z_{\mathrm{lens}}<0.9$, which brackets the redshift limits of the lenses in the Red-sequence Matched-filter Galaxy Catalog (RedMaGiC), described in Rozo et al. (in prep.), which selects red-sequence galaxies. The catalogue used here in what follows is limited to luminosity $L>L_{*}$, which results in approximately 30,000 lenses. To calculate statistical errors for the figures, we use three lens redshift bins: $0.2<z_{\mathrm{lens}}<0.4$, $0.4<z_{\mathrm{lens}}<0.6$, and $0.6<z_{\mathrm{lens}}<0.8$.

Figure \ref{fig:sigma_spec} shows the resulting $\Delta\langle \Sigma_{\mathrm{crit}}^{-1}\rangle/\langle \Sigma_{\mathrm{crit}}^{-1}\rangle_{\mathrm{spec}}$ for Test 2a. We find good agreement between the photometric estimates and the matched spectroscopic redshifts. \skynet and \tpz have biases that are nearly consistent with zero in all bins and lens redshifts, reaching levels comparable to the bootstrap errors over the spectroscopic sample at high redshift. The worst performing method, \bpz, has a bias that reaches only 15\% at the high redshift limit. For comparison, we include the statistical error on the magnitude of the tangential shear signal calculated via jackknife of the lens sample over the DES SV footprint. The weighted tangential shear $\gamma_t(\theta)$ enters into the calculation of $\Delta \Sigma$ linearly with $\Sigma_{\mathrm{crit}}^{-1}$. Except for \bpz, the bias for all methods is typically much less than this statistical error. We exclude Test 1 due to there being insufficient galaxies in the higher redshift bins to produce a ratio that is not dominated by noise, but have verified that in the lowest redshift bin, for example, there is negligible bias consistent with that shown in Fig. \ref{fig:sigma_spec} for Test 2a.

We repeat the same analysis for the full DES SV \ngmix\ shear catalogue in Fig. \ref{fig:sigma_sv}. The left panel shows $\langle \Sigma_{\mathrm{crit}}^{-1}\rangle$ as a function of lens redshift for each \photoz estimate, which agree well with each other. The differences are quantified in the right panels for each source redshift bin, where the fractional difference from the mean is shown. The spread in relative differences between the codes is within 5\% to that seen for the deep Test 2b in Fig. \ref{fig:sigma_spec}, which suggests that the bias shown in Fig.  \ref{fig:sigma_spec} is a good estimate of that expected in DES SV measurements of $\langle \Sigma_{\mathrm{crit}}^{-1}\rangle$.

\section{Conclusions}\label{sec:conclusions}

The Dark Energy Survey aims over five years of observations to combine the measurements of shapes and redshifts for hundreds of millions of galaxies to constrain cosmological parameters and to study the evolution and structure of dark energy and dark matter. The determination of accurate redshift distributions for these galaxies is one of the primary challenges for DES and for future weak lensing surveys, and may become the dominant systematic limitation in pursuing cosmology through precision weak lensing measurements. We have presented in this work an analysis of the resulting redshift distributions of galaxies with shape measurements from the pre-survey Science Verification data for DES (DES SV), and identified key challenges and obstacles in the pursuit of producing accurate redshift distributions for the main DES survey data releases at the level required to support ongoing DES weak lensing science.

We have compiled a set of more than 46,000 spectroscopic galaxies, which are matched in image depth and weighted to ensure even sampling of the \wls. These galaxies are split into training and validation samples, as well as an independent validation sample and a deep validation sample, the latter of which overlaps with the primary validation sample. The independent sample is taken from a separate spectroscopic field (VVDS-F14), while the deep sample is closer to the DES SV magnitude distribution. These spectroscopic samples are used as part of a larger test suite to verify and characterise the performance of the four photometric redshift codes compared in this paper: \annz, \bpz, \skynet, and \tpz. 

We identify challenges in producing photometric redshifts with the spectroscopic samples available to us and DES photometry, including learning the radial profile of the spectroscopic distributions in machine learning codes and mis-characterisation of the redshift in template-based approaches due to the limitations of our photometric bands and template colour coverage. This can result in artificial features in the photometric $n(z)$, which will bias any resulting analysis that depends on the photometric redshift distribution. We also discuss the challenge of compiling representative and complete spectroscopic training sets. However, we demonstrate that the potential bias in mean redshift due to spectroscopic incompleteness does not exceed the expected sample variance uncertainty in our presently available samples due to their small size.

In order to mitigate the potential issues associated with any given photometric redshift approach, we apply three independent methodologies: the first based on empirical spectroscopic data and utilising machine learning techniques; the second a modelling-based approach, comprising a template-fitting routine (\bpz) and a first-order correction of the associated model biases by image simulations (using BCC and UFig); and finally employing highly accurate empirical photometric redshifts from COSMOS, which have been selected to mimic our weak lensing sample. We find the mean redshift of the shear catalogue to be $z=0.72$. The variance in this mean and those of the three tomographic bins are consistent with Gaussian distribution of width 0.05. Therefore in the companion cosmology paper (DES et al. 2015), we marginalise over the photometric redshift calibration uncertainty using independent Gaussian priors of width 0.05 in each photometric redshift bin.

We propagate these \photoz uncertainties and biases through to measurements that are most relevant to weak lensing science, which is a necessary step to provide useful characterisations of \photoz biases for DES SV analysis papers. For each of the independent and deep weighted spectroscopic validation sets, we compare for each \photoz estimate the resulting measures of $\xi_+$ and the resulting constraints on $\sigma_8$, as well as resulting measurements of $\langle \Sigma_{\mathrm{crit}}^{-1}\rangle$. This provides us with direct estimates of expected biases on typical weak lensing measurements and cosmological parameters of interest, and allows us to validate methods of marginalising over \photoz biases.

We find that compared to the weighted spectroscopic validation sets, we should expect a level of bias for the fiducial \photoz estimates of less than about 10\% in $\xi_+$, which corresponds to a 1-$\sigma$ deviation or bias of $2-3\%$ in $\sigma_8$ for the fiducial \skynet method, given DES SV statistical power. We verify an approach to mitigate this bias by marginalising over bias parameters that shift the mean redshift of each tomographic bin, demonstrating that this is a sufficient approach to remove any bias in $\xi_+$ and $\sigma_8$. A similar analysis of $\langle \Sigma_{\mathrm{crit}}^{-1}\rangle$ finds a bias for the fiducial \photoz estimate that increases to approximately 5\% for the highest redshift lenses, but which is negligible for most lens redshifts.

Looking towards the future of the DES and beyond, weak lensing-oriented photo-z estimation will face a number of challenges. Firstly, in order to remain comparable to the expected statistical uncertainties in $5000~$deg$^2$ survey, the systematic uncertainties on the mean redshift within a given tomographic bin will need to be reduced from $\delta z \sim 0.05$ to an eventual level of $\delta z \sim 0.003$. Moreover, extracting the greatest amount of the information in the lensing signal will require the use of finer tomographic binning. 
Finally, the detailed topology of the $p(z)$ in a given tomographic bin will come under increasing scrutiny and marginalising over simple redshift bias parameters in the mean is unlikely to be sufficient in future cosmology analyses. Our testing metrics will need to be expanded to include those more sensitive to PDF information on a galaxy-by-galaxy basis \citep[e.g.][]{Bordoloi} in order to account for this shift in emphasis.

The methodologies employed to produce photo-zs can be improved upon by exploring better galaxy templates in modelling approaches to mitigate problems observed in this work, and the incorporation of galaxy information beyond magnitude and colour may be key to breaking degeneracies in the machine learning PDFs. Coupled with algorithmic improvements is the increasing availability of data. For instance, the year 1 DES survey data cover further key spectroscopic fields in Stripe 82, BOSS, DEEP2 and Wigglez. Wide field spectroscopic fields, even those biased towards the brightest objects, open up new possibilities in the form of cross-correlation analyses \citep{Newman08}. Meanwhile, further exquisite {\em photometric} fields will also be covered and should allow us to conduct comparisons similar to the one we performed with COSMOS in this work, but with reduced sample variance concerns. Despite these foreseen advances in weak lensing photo-z techniques, there still remains the separate issue of {\em validating} the derived redshifts. To be fully confident in both the redshifts and the estimated uncertainties that we find with the various photo-z techniques, the need for additional deep, but highly complete, spectroscopy is unavoidable.

\section*{Acknowledgements}

We are grateful for the extraordinary contributions of our CTIO colleagues and the DECam 
Construction, Commissioning and Science Verification teams in achieving the excellent 
instrument and telescope conditions that have made this work possible. The success of this 
project also relies critically on the expertise and dedication of the DES Data Management group.

MT, SB, NM, and JZ acknowledge support from the European Research Council in the form of a Starting Grant with number 240672.
DG  acknowledges the support by the DFG Cluster of Excellence ÓOrigin and Structure of the UniverseÓ

Funding for the DES Projects has been provided by the U.S. Department of Energy, the U.S. National Science 
Foundation, the Ministry of Science and Education of Spain, the Science and Technology Facilities Council of 
the United Kingdom, the Higher Education Funding Council for England, the National Center for Supercomputing 
Applications at the University of Illinois at Urbana-Champaign, the Kavli Institute of Cosmological Physics 
at the University of Chicago, the Center for Cosmology and Astro-Particle Physics at the Ohio State University,
the Mitchell Institute for Fundamental Physics and Astronomy at Texas A\&M University, Financiadora de 
Estudos e Projetos, Funda{\c c}{\~a}o Carlos Chagas Filho de Amparo {\`a} Pesquisa do Estado do Rio de 
Janeiro, Conselho Nacional de Desenvolvimento Cient{\'i}fico e Tecnol{\'o}gico and the Minist{\'e}rio da 
Ci{\^e}ncia e Tecnologia, the Deutsche Forschungsgemeinschaft and the Collaborating Institutions in the 
Dark Energy Survey. 

Gangkofner acknowledges the support by the DFG Cluster of Excellence ÓOrigin and Structure of the UniverseÓ

The DES data management system is supported by the National Science Foundation under Grant Number 
AST-1138766. The DES participants from Spanish institutions are partially supported by MINECO under 
grants AYA2012-39559, ESP2013-48274, FPA2013-47986, and Centro de Excelencia Severo Ochoa 
SEV-2012-0234, some of which include ERDF funds from the European Union.

The Collaborating Institutions are Argonne National Laboratory, the University of California at Santa Cruz, 
the University of Cambridge, Centro de Investigaciones Energeticas, Medioambientales y Tecnologicas-Madrid, 
the University of Chicago, University College London, the DES-Brazil Consortium, the Eidgen{\"o}ssische 
Technische Hochschule (ETH) Z{\"u}rich, Fermi National Accelerator Laboratory,
the University of Edinburgh, 
the University of Illinois at Urbana-Champaign, the Institut de Ci\`encies de l'Espai (IEEC/CSIC), 
the Institut de F\'{\i}sica d'Altes Energies, Lawrence Berkeley National Laboratory, the Ludwig-Maximilians 
Universit{\"a}t and the associated Excellence Cluster Universe, the University of Michigan, the National Optical 
Astronomy Observatory, the University of Nottingham, The Ohio State University, the University of Pennsylvania, 
the University of Portsmouth, SLAC National Accelerator Laboratory, Stanford University, the University of 
Sussex, and Texas A\&M University.

Based in part on observations taken at the Australian Astronomical Observatory under program A/2013B/012.

Parts of this research were conducted by the Australian Research Council Centre of Excellence for All-sky Astrophysics (CAASTRO), through project number CE110001020.

This work was supported in part by grants $200021\_14944$ and $200021\_143906$ from the Swiss National Science Foundation.

Funding for SDSS-III has been provided by the Alfred P. Sloan Foundation, the Participating Institutions, the National Science Foundation, and the U.S. Department of Energy Office of Science. The SDSS-III web site is http://www.sdss3.org/.

SDSS-III is managed by the Astrophysical Research Consortium for the Participating Institutions of the SDSS-III Collaboration including the University of Arizona, the Brazilian Participation Group, Brookhaven National Laboratory, Carnegie Mellon University, University of Florida, the French Participation Group, the German Participation Group, Harvard University, the Instituto de Astrofisica de Canarias, the Michigan State/Notre Dame/JINA Participation Group, Johns Hopkins University, Lawrence Berkeley National Laboratory, Max Planck Institute for Astrophysics, Max Planck Institute for Extraterrestrial Physics, New Mexico State University, New York University, Ohio State University, Pennsylvania State University, University of Portsmouth, Princeton University, the Spanish Participation Group, University of Tokyo, University of Utah, Vanderbilt University, University of Virginia, University of Washington, and Yale University.

Based on observations made with ESO Telescopes at the La Silla Paranal Observatory under programme ID 179.A-2004.
Based on observations made with ESO Telescopes at the La Silla Paranal Observatory under programme ID 177.A-3016.

This paper is Fermilab publication
FERMILAB-PUB-15-306 and DES publication
DES2015-0060. This paper has gone through internal
review by the DES collaboration.

~\\

\begin{appendices}
\section{Details of matched spectroscopic sample}\label{section:app_a}
In this appendix, we note all the quality flags that are used in the \msc and their meaning.

\begin{description}
\item[$\cdot$ 2dFGRS:] All galaxies with flags 3, 4 or 5,  all of these are considered to be reliable redshifts \citep{2df}.
\item[$\cdot$ ACES:] All  galaxies with flags 3 or 4, these are labeled as secure and very secure redshifts \citep{aces}.
\item[$\cdot$ ATLAS:] This survey \citep{atlas} has no quality flags, all objects classified as galaxies where kept.
\item[$\cdot$ OzDES:] All galaxies with quality flag 4,  galaxies with this flag are expected to have the correct redshift more than 99\% of the time \citep{ozdes}
\item[$\cdot$ ELG Cosmos:] All galaxies with quality flags 3 or 4, these correspond to clear single line redshift identification and a secure redshift respectively \citep{ELG_cosmos}.
\item[$\cdot$ GAMA:] All galaxies with quality flag 4, these are labelled as certain redshifts \citep{gama}.
\item[$\cdot$ PanSTARRS AAOmega:] All galaxies with quality flag 3 or 4, galaxies with these flags are expected to have the correct redshift more than 95\% or 99\% of the time, respectively. \citep{PS1,PS2,PS3}.
\item[$\cdot$ PanSTARRS MMT:] All galaxies with quality flag 3 or 4, these are labelled as probably and as certain redshifts \citep{PS1, PS2, PS3}.
\item[$\cdot$ SDSS DR10:] All galaxies with quality flag 0, this are labelled as reliable \citep{dr10}.
\item[$\cdot$ SNLS AAOmega:] All galaxies with quality flag 4 and 5, these are labelled as reliable and reliable with more the 3 clearly visible features \citep{snls_aaomega}.
\item[$\cdot$ SNLS] All galaxies with quality flag 1 and 2, these are labelled as reliable based on several strong detected features and on one clearly detected feature, usually [OII]  \citep{ozdes2}.
\item[$\cdot$ UDS:] All galaxies observed with VIMOS that have quality flags 3 and 4, these are labelled as secure.
All galaxies observed with FORS2 that have quality flags A, B or B* where A and B  is labeled as secure and B* is labeled as reliable.
See \cite{UDS1,UDS2} for more information.
\item[$\cdot$ VIPERS:] All galaxies that have flags 3 and 4, these are labeled as reliable \citep{vipers}. 
\item[$\cdot$ Zcosmos:] All galaxies that have flags 3 and 4, these are labeled as secure and very secure redshifts \citep{zcosmos}. 
\item[$\cdot$ VVDS:]  All galaxies that have flags 3 and 4, these are labeled as secure and very secure redshifts \citep{vvdsw, vvdsd}.
\end{description}

\section{Self-Selection tomographic analysis}
\label{sec:app_tomo}

\begin{figure*}
\begin{center}
\includegraphics[scale=0.75]{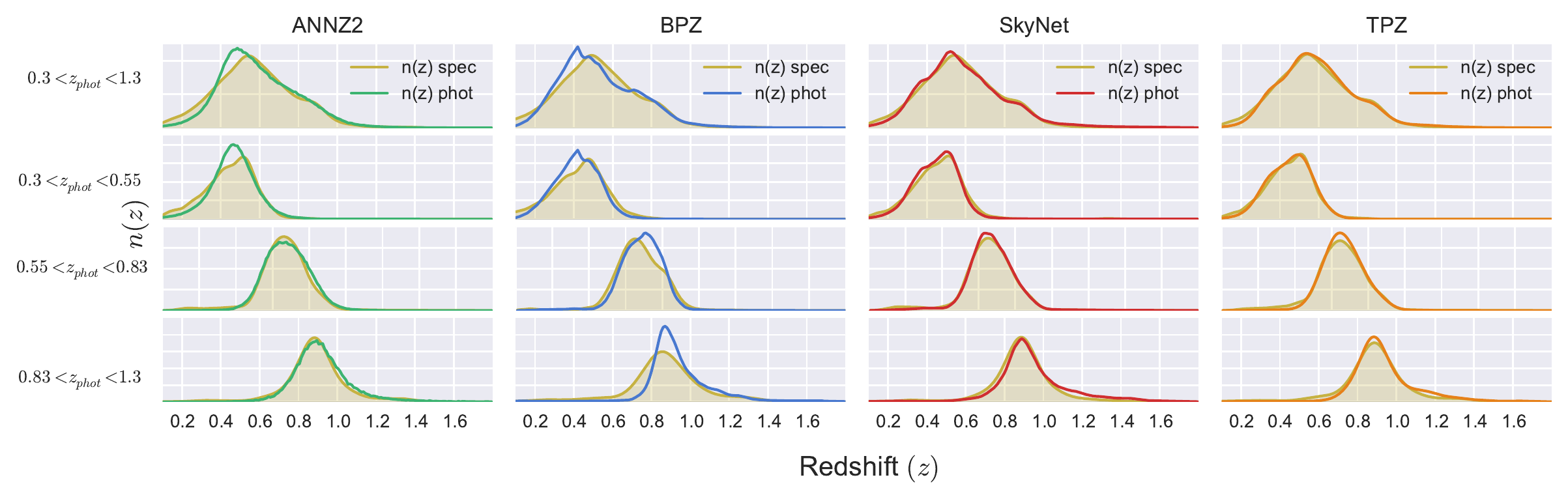}
\end{center}
\caption[]{The weighted spectroscopic redshift distribution $n(z)$ (shaded area) compared to the estimates of the four codes for the Test 1 (VVDS-F14) galaxies. Unlike in Sec. \ref{sec:tomo}, all codes assign galaxies to tomographic bins according to their own mean PDF estimates, hence the objects in each bin differ for each panel in the plot.  
\label{fig:test1_nz}}
\end{figure*}

\begin{figure*}
\begin{center}
\includegraphics[scale=0.75]{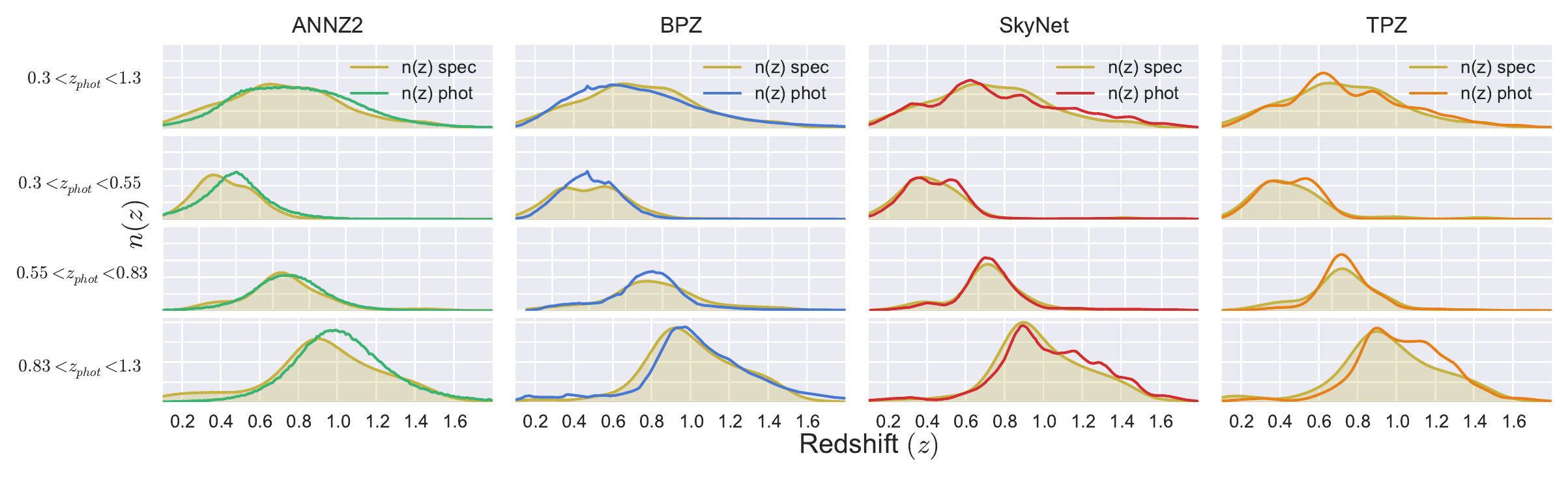}
\end{center}
\caption[]{The weighted spectroscopic redshift distribution $n(z)$ (shaded area) compared to the estimates of the four codes in Test 2a (VVDS-Deep galaxies in the validation set). 
Unlike in Sec. \ref{sec:tomo}, all codes assign galaxies to tomographic bins according to their own mean PDF estimates, hence the objects in each bin differ for each panel in the plot.  
\label{fig:test2a_nz}}
\end{figure*}

\begin{figure*}
\begin{center}
\includegraphics[scale=0.75]{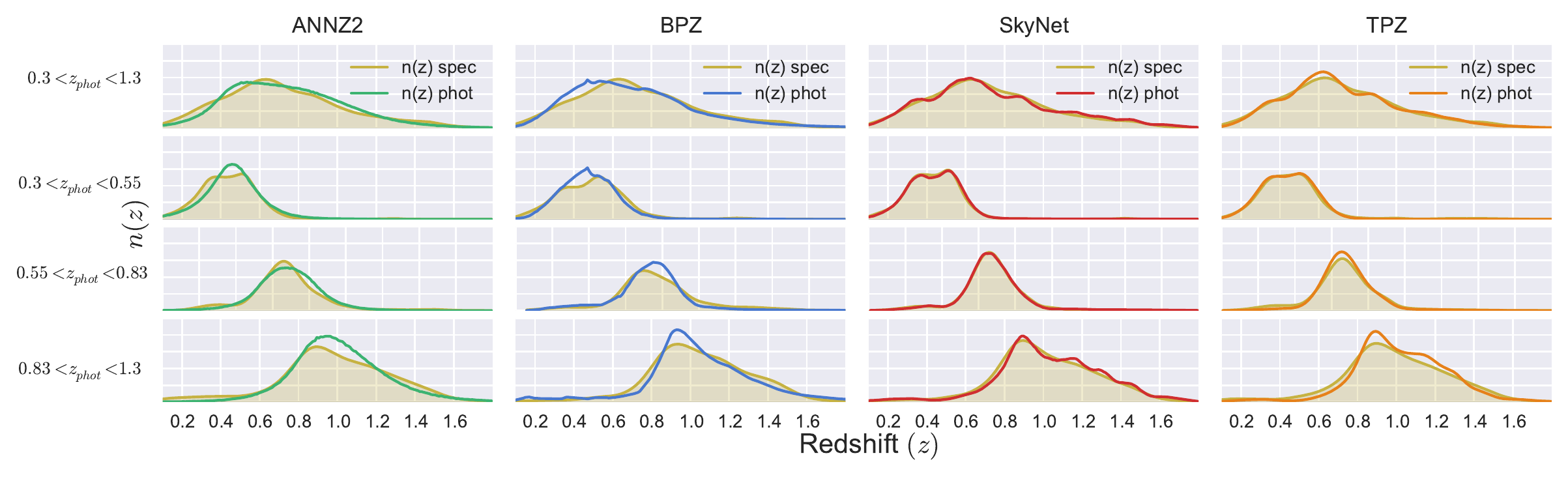}
\end{center}
\caption[]{The weighted spectroscopic redshift distribution $n(z)$ (shaded area) compared to the estimates of the four codes in Test 2b (Full validation set).
Unlike in Sec. \ref{sec:tomo}, all codes assign galaxies to tomographic bins according to their own mean PDF estimates, hence the objects in each bin differ for each panel in the plot.  
\label{fig:test2b_nz}}
\end{figure*}

We repeat here Tests 1, 2a, and 2b, but now allow each code to assign a galaxy to each redshift bin based on it's own estimate of the mean PDF instead of that of \skynet\ as was done in Sec. \ref{sec:tomo}. 
Figs. \ref{fig:test1_nz}, \ref{fig:test2a_nz}, and \ref{fig:test2b_nz} show the performance of the four methods. Table \ref{table:test12a2bown} 
shows the offsets of the mean of the redshift estimated distributions with respect to the weighted spectroscopic distribution.
There is not a clear benefit to enforcing separate tomographic binning based on each photo-z method and repeating the analysis pipelines in the companion papers for DES SV, as some methods perform better and others worse when using the fiducial \skynet binning.

\begin{table}
\begin{center}
\begin{tabular}{ccrrrr}
\toprule
& z range& \annz &  \bpz &    \skynet &    \tpz \\
\midrule
\multirow{3}{*}{Test 1} &$0.30 - 0.55 $   & 0.017  &  -0.005  &  0.003  &  0.004  \\  
&$0.55 - 0.83 $                           & 0.018  &  0.01    &  0.017  &  0.016  \\ 
&$0.83 - 1.30 $                           & 0.032  &  0.077   &  0.063  &  0.050  \\
\midrule
\multirow{3}{*}{Test 2a} & $0.30 - 0.55 $    &  0.049  &   0.002  &  0.027  &  -0.013  \\ 
&$0.55 - 0.83 $                              &  0.015  &  -0.025  &  0.034  &   0.031  \\  
&$0.83 - 1.30 $                              &  0.086  &   0.046  &  0.044  &   0.069  \\  
\midrule
\multirow{3}{*}{Test 2b} &$0.30 - 0.55 $      &  0.015  &  -0.015  &  0.012  &  -0.020  \\
&$0.55 - 0.83 $                               &  0.011  &  -0.027  &  0.013  &   0.008  \\  
&$0.83 - 1.30 $                               &  0.025  &   0.007  &  0.022  &   0.028  \\ 
\bottomrule
\end{tabular}
\end{center}
\caption{The bias $(\langle z_{phot}\rangle-\langle z_{spec}\rangle)$ between the photometric redshift estimates and the true spectroscopic distribution in Test 1 (independent), Test 2a (VVDS-Deep), and Test 2b (Full validation set) when the codes each assign their own binning to the galaxies.}
\label{table:test12a2bown}
\end{table}

\section{Photo-z methods}\label{app:methods}
\subsection{ANNZ2}

\annz\ \citep{annz2} \footnote{\url{https://github.com/IftachSadeh/ANNZ}} is an updated version of the neural network code ANNz (\citealt{collisterlahav}). 
\annz\ differs from its previous version by incorporating several additional machine learning methods beyond Artificial Neural Networks (ANNs), such as Boosted Decision Trees (BDTs) and $k$-Nearest Neighbours (KNN) algorithms. 
These are implemented in the TMVA package (\citealt{tmva})\footnote{TMVA is a part of the ROOT C++ software framework (\cite{root})}.

For the 100 ANNs run on the spectroscopic training set, we randomly varied: the number of nodes in each layer, the number of training cycles, the usage of the so-called \emph{Bayesian regulator}, that reduces the risk of over-training, the type of activation function, the type of variable transformation performed before training (such as normalisation and PCA transformation), the number of subsequent convergence tests which have to fail to consider the training complete, and the initial random seed.  After training is complete,  the performance of each method is quantified through an optimisation process, which leads to a single nominal \photoz estimator for \annz. The entire collection of solutions is used in order to derive a $p(z)$, constructed in two steps.
First, each solution is folded with an error distribution, which is derived using the KNN error estimation method of \citealt{oyaizu}.
The ensemble of solutions is then combined using an optimised weighting scheme. This methodology allows us to take into account both the intrinsic errors on the input parameters for a given method, and the uncertainty on the method itself. The methodology described above is what is called "randomised regression".
Another important feature implemented in ANNz2 is the weighting method (\citealt{lima}). It is therefore possible to give in input a reference sample and re-weight the training set to make its relevant variables distributions more representative of the former, this was technique was applied in this work.

\subsection{BPZ}
The \bpz\ (Bayesian Photometric Redshifts) \photoz code \citep{Benitez2000,Coe2006} is a model fitting code 
that fits galaxy templates to the measured photometry and its associated errors.
\bpz\ calculates the likelihood of the galaxy for the best fitting template, 
which then, using Bayes theorem, is combined with a prior to produce the likelihood.
The prior represents our previous knowledge of the redshift
and spectral type distributions of the sample in the analysis.

\begin{itemize}
\item Templates: We use the eight spectral templates that \bpz
carries by default based on \cite{Coleman1980, kinney1996}, and add two more interpolated templates
between each pair of them by setting the input parameter
\textsc{INTERP=8} (option by default). 
\item Prior: We explicitly calibrate the prior in each test by fitting
the empirical function $\Pi(z, t \mid m_0)$ proposed in \cite{Benitez2000}
to the weighted training set, although we note that using the weighted or unweighted training set to
get the prior had a negligible effect on \photoz performance.   
\end{itemize}

\subsection{SkyNet}
\skynet\ \citep{Graff2013} is a neural network algorithm that uses a 2nd order method based on a conjugate gradient algorithm to find the optimal weights 
of the network. \skynet\ classifies galaxies in classes, in this case
redshift bins, where the last layer is a softmax transformation that is able to estimate the probability that an
object belongs to a certain class (or bin) \citep{Bonnett2015}. 
The number of classes is the redshift bin resolution of the pdf.
In this work \skynet\ is run slightly different than in \cite{sanchez2014dessva1photoz,Bonnett2015}.
\skynet\ is run 10 times with the same network configuration but with a slightly shifted binning each time. 
We train with a nominal bin width of $\Delta z = 0.09$ -- these are referred to as the broad bins.
The broad bins are then slightly shifted by $\delta = 0.009$ every training run so that $\Delta z$ is sampled in 10 locations, leading to a overall sampling of  $\delta z = 0.009$. This produces 200 bins between
$z = 0.005$ and $z = 1.8$.
After the 10 networks have been trained, the pdf values at $z_i$ are taken to be the average of all the broad bins that $z_i$ lies within.
This means that the \skynet\ photometric redshifts have an intrinsic smoothing built into them.
All the networks have the same architecture, 3 layers with 16, 14, and 20 nodes per layer and a $tanh$ activation function.
The features fed to the network are the \magauto\ $i, r$ and all possible colour combinations of the four bands.
In this work we make use of the python wrapper \textsc{pySkyNet} \footnote{\url{http://pyskynet.readthedocs.org/}} of the \skynet\ library.
    
\subsection{TPZ}
\textsc{tpz}\footnote{\url{http://lcdm.astro.illinois.edu/research/TPZ.html}} \citep{Carrasco2013} is a machine learning algorithm that uses prediction trees and random forest techniques to produce robust photometric redshift PDFs. 
Prediction trees are built by asking a series of questions that recursively split the input data taken from the spectroscopic
sample into two branches, until a terminal leaf is created that meets the stopping criterion. 
The method by which the data are divided is chosen to be the one with highest information
gain among the random subsample of features chosen at every point. 
This produces less correlated trees that act as weak learners that can be combined into a strong predictor.
All objects in a terminal leaf node represents a specific subsample of the entire data with similar properties.
Additional data is created before the trees are constructed by perturbing the data using their magnitude errors -- this is sometimes referred to as a parametric bootstrap.
In this work 200 trees were created whose results were aggregated to construct each individual PDF. 
For the application to DES SV data, we have used {\it griz} \magauto\ magnitudes together with all the corresponding colours and their associated errors. We discretised the redshift space into 100 bins up to $z=1.8$ and adopted a smoothing scale of $5$ times the bin size.
 
\section{PRIMUS, an example of extreme selection  effects.}\label{app:primus}

\begin{figure}
\begin{center}
\includegraphics[width=\columnwidth]{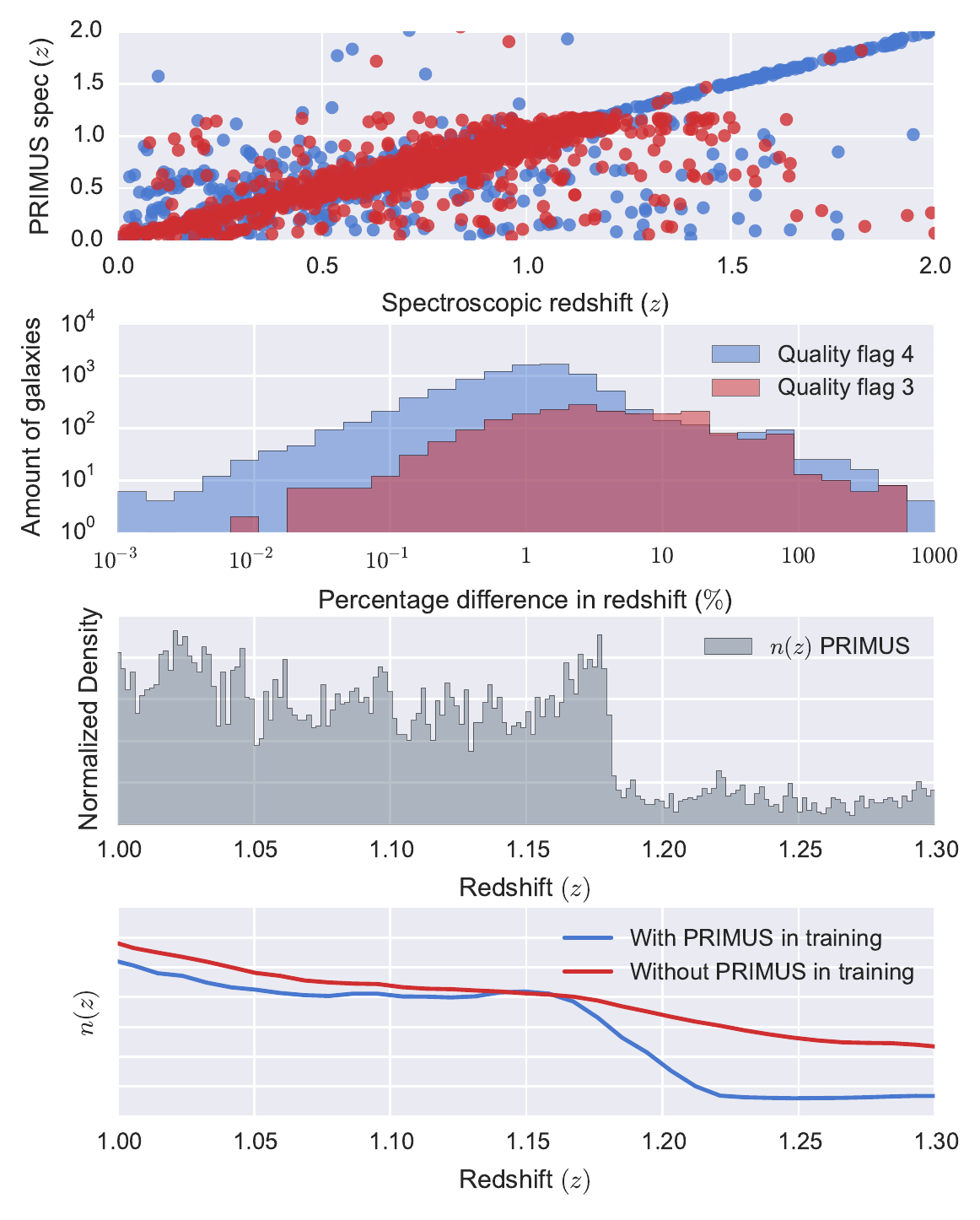}
\end{center}
\caption[]{An analysis of challenges related to the use of PRIMUS spectroscopic redshifts as part of the DES SV training or validation samples. Top panel: PRIMUS redshift vs the matched spectroscopic redshift from higher resolution instruments. The blue dots are the highest quality flag 4, while the red dots are the second highest quality flag 3. Second panel:
The fractional difference of the redshifts between PRIMUS and the other surveys. Third panel: The spectroscopic redshift distribution of PRIMUS galaxies between $1.0 < z < 1.3$. Around $\thicksim$ 1.2, there is a large drop in the spectroscopic redshift distribution due to fact the galaxies have a maximum fitting redshift of $z=1.2$, while AGN are fit up to $z=5.0$. 
Bottom panel: The effect of this drop in PRIMUS $n(z)$ on the final estimation of the $n(z)$ for the DES SV shear catalogue. Shown are two examples of including or excluding the PRIMUS galaxies using \skynet, where the feature at $z=1.2$ is clearly imprinted on the $n(z)$ of the \wls\ when PRIMUS galaxies are included in the training.
\label{fig:prim1}}
\end{figure}

In building the spectroscopic training and validation samples, we have excluded any galaxies from the PRIMUS survey \citep{PRIMUS}. 
Here we will discuss some of the complications of using PRIMUS galaxies as part of the training or validation samples.
PRIMUS is a spectroscopic survey covering a total of $9.1$ deg$^2$ containing 185,105 galaxies, of which we have matched 88,040 galaxies that have DES SV photometry within 1.5 arcseconds only using the two highest PRIMUS quality flags 4 and 3.
The PRIMUS redshifts are obtained by fitting low resolution spectra and any matched photometry to an empirical library of spectra based on the AGES spectra \citep{AGES}.
The PRIMUS redshifts have two peculiarities, the first being that a non-negligible amount of galaxies have a different redshift when compared to
objects with spectra from higher resolution instruments. 
\cite{PRIMUS} estimate $\sigma_{\delta z/(1+z)} = 0.005$ and $0.022$ for quality flags 4 and 3, while we find $0.004$ and $0.010$ for all the matched objects within the DES survey.
The top two panels of Fig. \ref{fig:prim1} show this comparison of the PRIMUS spectroscopic redshifts with matched spectroscopic redshifts from higher resolution instruments. This leads us to consider the unresolved question of how robust ML and other calibration methods are to incorrect spectra, which is not a question that we attempt to answer in this work, but one for which there has been some work in general in the ML literature (\eg \cite{ml_noise1,2014MNRAS.444..129C}).

The second PRIMUS feature that is important for photometric redshift estimation is the fact that galaxies are only fit up to $z=1.2$.
The cut at $z=1.2$ is effectively a selection effect and hence, one must take care when using PRIMUS to train.
To illustrate this, consider a galaxy at $z=1.2$ observed by PRIMUS and DES for which we want to estimate the $p(z)$. 
In the idealised case of a Gaussian pdf, the mean would be located around
$z=1.2$ and there would be tails in the $p(z)$ extending to lower and higher redshift. 
Given that there are no galaxies beyond $z=1.2$ in PRIMUS, none of the ML methods will be able to learn that some probability should extend beyond $z=1.2$. 
Even when assessing how well a template fitting method performs, the lack of spectra beyond $z=1.2$ may lead one to believe the performance is poor.
These features are demonstrated in the bottom two panels of Fig. \ref{fig:prim1}.
In the bottom panel of Fig. \ref{fig:prim1}, we provide a real example of the difference on the reconstructed $n(z)$ for the \wls\ around $z=1.2$ when trained with and without PRIMUS.
Though this is an extreme case of a selection effect imprinting itself on the reconstructed $n(z)$, it is possible that similar, more subtle effects persist in the ML photometric redshift estimates. There are a large number of PRIMUS spectra, however, and careful efforts should be made to find ways to utilise these in the future.

\end{appendices}

\setlength{\bibhang}{2.0em}
\setlength\labelwidth{0.0em}
\bibliographystyle{mn2e_good}
\bibliography{paper}

~
\newline
$^{1}$Institut de F\'{\i}sica d'Altes Energies, Universitat Aut\`onoma de Barcelona, E-08193 Bellaterra, Barcelona, Spain\\
$^{2}$Jodrell Bank Center for Astrophysics, School of Physics and Astronomy, University of Manchester, Oxford Road, Manchester, M13 9PL, UK\\
$^{3}$Department of Physics, ETH Zurich, Wolfgang-Pauli-Strasse 16, CH-8093 Zurich, Switzerland\\
$^{4}$Department of Physics \& Astronomy, University College London, Gower Street, London, WC1E 6BT, UK\\
$^{5}$Department of Physics, Stanford University, 382 Via Pueblo Mall, Stanford, CA 94305, USA\\
$^{6}$Kavli Institute for Particle Astrophysics \& Cosmology, P. O. Box 2450, Stanford University, Stanford, CA 94305, USA\\
$^{7}$Department of Physics and Astronomy, University of Pennsylvania, Philadelphia, PA 19104, USA\\
$^{9}$Department of Astronomy, University of Illinois, 1002 W. Green Street, Urbana, IL 61801, USA\\
$^{10}$National Center for Supercomputing Applications, 1205 West Clark St., Urbana, IL 61801, USA\\
$^{11}$Institut de Ci\`encies de l'Espai, IEEC-CSIC, Campus UAB, Carrer de Can Magrans, s/n,  08193 Bellaterra, Barcelona, Spain\\
$^{12}$Jet Propulsion Laboratory, California Institute of Technology, 4800 Oak Grove Dr., Pasadena, CA 91109, USA\\
$^{13}$Fermi National Accelerator Laboratory, P. O. Box 500, Batavia, IL 60510, USA\\
$^{14}$Kavli Institute for Cosmological Physics, University of Chicago, Chicago, IL 60637, USA\\
$^{15}$Argonne National Laboratory, 9700 South Cass Avenue, Lemont, IL 60439, USA\\
$^{16}$Australian Astronomical Observatory, North Ryde, NSW 2113, Australia\\
$^{17}$Instituci\'o Catalana de Recerca i Estudis Avan\c{c}ats, E-08010 Barcelona, Spain\\
$^{18}$Department of Physics, University of Arizona, Tucson, AZ 85721, USA\\
$^{19}$SLAC National Accelerator Laboratory, Menlo Park, CA 94025, USA\\
$^{20}$Centre for Astrophysics \& Supercomputing, Swinburne University of Technology, Victoria 3122, Australia\\
$^{21}$Cerro Tololo Inter-American Observatory, National Optical Astronomy Observatory, Casilla 603, La Serena, Chile\\
$^{22}$Department of Astrophysical Sciences, Princeton University, Peyton Hall, Princeton, NJ 08544, USA\\
$^{23}$Institute of Astronomy, University of Cambridge, Madingley Road, Cambridge CB3 0HA, UK\\
$^{24}$Kavli Institute for Cosmology, University of Cambridge, Madingley Road, Cambridge CB3 0HA, UK\\
$^{25}$CNRS, UMR 7095, Institut d'Astrophysique de Paris, F-75014, Paris, France\\
$^{26}$Sorbonne Universit\'es, UPMC Univ Paris 06, UMR 7095, Institut d'Astrophysique de Paris, F-75014, Paris, France\\
$^{27}$Institute of Cosmology \& Gravitation, University of Portsmouth, Portsmouth, PO1 3FX, UK\\
$^{28}$Laborat\'orio Interinstitucional de e-Astronomia - LIneA, Rua Gal. Jos\'e Cristino 77, Rio de Janeiro, RJ - 20921-400, Brazil\\
$^{29}$Observat\'orio Nacional, Rua Gal. Jos\'e Cristino 77, Rio de Janeiro, RJ - 20921-400, Brazil\\
$^{30}$George P. and Cynthia Woods Mitchell Institute for Fundamental Physics and Astronomy, and Department of Physics and Astronomy, Texas A\&M University, College Station, TX 77843,  USA\\
$^{31}$Department of Physics, Ludwig-Maximilians-Universit\"at, Scheinerstr. 1, 81679 M\"unchen, Germany\\
$^{32}$Excellence Cluster Universe, Boltzmannstr.\ 2, 85748 Garching, Germany\\
$^{33}$Universit\"ats-Sternwarte, Fakult\"at f\"ur Physik, Ludwig-Maximilians Universit\"at M\"unchen, Scheinerstr. 1, 81679 M\"unchen, Germany\\
$^{34}$Department of Physics, University of Michigan, Ann Arbor, MI 48109, USA\\
$^{35}$Max Planck Institute for Extraterrestrial Physics, Giessenbachstrasse, 85748 Garching, Germany\\
$^{36}$Center for Cosmology and Astro-Particle Physics, The Ohio State University, Columbus, OH 43210, USA\\
$^{37}$Department of Physics, The Ohio State University, Columbus, OH 43210, USA\\
$^{38}$Lawrence Berkeley National Laboratory, 1 Cyclotron Road, Berkeley, CA 94720, USA\\
$^{39}$Department of Astronomy, The Ohio State University, Columbus, OH 43210, USA\\
$^{40}$Department of Astronomy, University of Michigan, Ann Arbor, MI 48109, USA\\
$^{41}$Department of Physics and Astronomy, Pevensey Building, University of Sussex, Brighton, BN1 9QH, UK\\
$^{42}$Centro de Investigaciones Energ\'eticas, Medioambientales y Tecnol\'ogicas (CIEMAT), Madrid, Spain\\
$^{43}$Instituto de F\'\i sica, UFRGS, Caixa Postal 15051, Porto Alegre, RS - 91501-970, Brazil\\
$^{44}$Department of Physics, University of Illinois, 1110 W. Green St., Urbana, IL 61801, USA\\
$^{45}$Departamento de F\'\i sica Matem\' atica, Instituto de F\'\i sica, Universidade de S\~ao Paulo,CP 66318, CEP 05314-970 S\~ao Paulo, SP Brazil \\
$^{46}$SEPnet, South East Physics Network, (www.sepnet.ac.uk) \\
$^{47}$School of Mathematics and Physics, University of Queensland, QLD 4072, Australia \\
$^{48}$Research School of Astronomy and Astrophysics, Australian National University,  Canberra, ACT 2611, Australia.\\
$^{49}$Centre for Astrophysics \& Supercomputing, Swinburne University of Technology, Victoria 3122, Australia \\
$^{50}$Brookhaven National Laboratory, Bldg 510, Upton, NY 11973, USA\\
$^{51}$Excellence Cluster Universe, Boltzmannstr.\ 2, 85748 Garching, Germany \\
$^{52}$Faculty of Physics, Ludwig-Maximilians University, Scheinerstr. 1, 81679 Munich, Germany
\end{document}